\documentclass[sigconf,10pt,noacm]{acmart}
\renewcommand\footnotetextcopyrightpermission[1]{} 
\usepackage{tikz}
\usepackage{amsmath}

\usepackage{filecontents}
\usepackage{amsbsy}
\usepackage{algorithmic}
\usepackage{balance}
\usepackage{caption}
\usepackage{gensymb}
\usepackage{graphicx}
\usepackage{multirow}
\usepackage{subfigure}
\usepackage{textcomp}
\usepackage{url}
\usepackage[english]{babel}
\usepackage{blindtext}
\usepackage{colortbl}

\usepackage{tikz}
\usepackage{amsmath,amssymb,amsfonts,mathrsfs}
\usepackage{latexsym}
\usepackage{algorithmic}
\usepackage{graphicx}
\usepackage{subfigure}
\usepackage{gensymb}
\usepackage{mathtools}
\usepackage{url}
\usepackage{balance}
\usepackage{xspace}
\usepackage{multirow}
\usepackage{tabularx}
\usepackage{makecell}
\usepackage{textcomp}
\usepackage{caption}
\usepackage{enumitem}
\usepackage{booktabs}
\usepackage{pifont}
\usepackage[vlined,ruled,linesnumbered]{algorithm2e}
\usepackage{pdfpages}
\usepackage{setspace}

\long\def\comment#1{}
\long\def\delete#1{}

\newcommand{\sysname}{\textsc{MetaSeq}\xspace}

\newcommand{\sssec}[1]{\vspace*{0.025in}\noindent\textbf{#1} }

\newenvironment{packeditemize}{\begin{list}{$\bullet$}{\setlength{\itemsep}{0.5pt}\addtolength{\labelwidth}{-1pt}\setlength{\leftmargin}{\labelwidth}\setlength{\listparindent}{\parindent}\setlength{\parsep}{1pt}\setlength{\topsep}{0pt}}}{\end{list}}

\begin{document}



\title{Physics-Guided Sequence-Based Generative Framework for Acoustic Metamaterial Inverse Design}

\author{Yijie Li}
\affiliation{%
	\institution{National University of Singapore, Singapore}
	\country{}
}
\email{yijieli@nus.edu.sg}

\author{Jiahao Xu}
\affiliation{%
	\institution{National University of Singapore, Singapore}
	\country{}
}
\email{e1583425@u.nus.edu}

\author{Ching-Chih Tsao}
\affiliation{%
	\institution{National University of Singapore, Singapore}
	\country{}
}
\email{cctsao@u.nus.edu}

\author{Lili Qiu}
\affiliation{%
	\institution{UT Austin}
	\country{USA}
}
\email{lili.qiu.cs@gmail.com}

\author{Jingxian Wang}
\affiliation{%
	\institution{National University of Singapore, Singapore}
	\country{}
}
\email{wang@nus.edu.sg}


\begin{abstract}
Acoustic metamaterial (AMM) inverse design is particularly challenging for broadband target responses due to acoustic dispersion: a structure that matches the desired response at one frequency may deviate at others, and modifying geometry to improve one sub-band often perturbs neighboring sub-bands. Yet existing broadband inverse-design approaches are either constrained by predefined templates, or rely on image representations that fail to preserve the geometric precision and structural connectivity required by acoustic structures. We present \sysname, a physics-guided, sequence-based generative framework for acoustic metamaterial inverse design. At its core, \sysname introduces a  language that represents each AMM as a structured sequence, rather than as a pixel grid or fixed template. This representation preserves  precise geometry, explicitly encodes connectivity, and casts inverse design as a sequence-to-sequence task from target response to structure sequence. \sysname further constructs a balanced, high-fidelity dataset with efficient calibration and  complexity-based sampling. To address the one-to-many nature of inverse design, \sysname combines supervised pretraining with reinforcement learning fine-tuning guided by a physics-based solver and validity checker. Extensive evaluations against COMSOL and five baselines show that \sysname reduces response error by 45\% over the best baseline.
\end{abstract}
\maketitle

\section{Introduction}
Acoustic metamaterials (AMMs)~\cite{fok2008acoustic} are artificially engineered sub-wavelength acoustic units, each composed of components such as cavities and  ducts that shape acoustic responses. These  structures serve as  building blocks for acoustic functions such as sound absorption~\cite{Ma2014NatMater,LiAssouar2016APL,li2023sound,yang2022development,guo2023phase,ryoo2022broadband}, and wave manipulation~\cite{zhang2021remote,ning2025portable,lu2023achromatic,fu2024adaptive,li2024mudis}, and can also be arranged into larger acoustic metasurfaces~\cite{Assouar2018NRM}.


This work focuses on the inverse design of AMM units, where the goal is to generate a structure\footnote{An AMM unit is an individual acoustic structure that may function either as a standalone device or as part of a larger acoustic metasurface. This paper focuses on AMM unit design rather than metasurface-level design.} that realizes a target response, i.e.,  desired phase and amplitude. This problem becomes substantially challenging for broadband targets~\cite{guo2023phase,ryoo2022broadband,lu2023achromatic,chen2018broadband}, which require a desired response over a frequency range, e.g., high sound absorption across a band from 100 Hz to 1 kHz. The  challenge stems from  the dispersive nature of acoustics~\cite{bruneau2013fundamentals}: the same geometry can exhibit substantially different responses across 
frequencies. As a result, a structure matching the target at one frequency may deviates at others, and modifying the geometry to improve one sub-band often alters the response in neighboring sub-bands, making broadband design challenging.

\begin{figure}[t]
	\centering
	\includegraphics[width=0.8\linewidth]{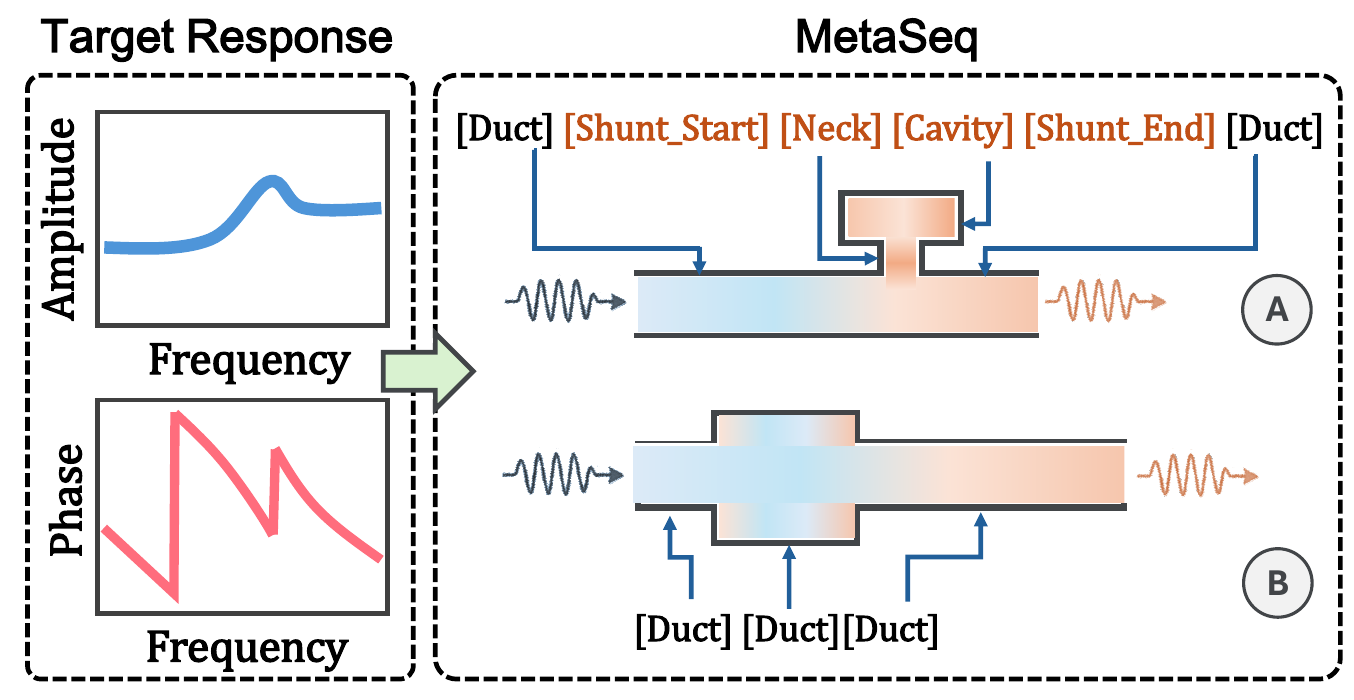}
	\vspace{-10pt}
	\caption{\sysname casts  acoustic metamaterial inverse design as  generating a sequence representation of an AMM from a target acoustic response.} 
	\label{fig:fig1}
	\vspace{-15pt}
\end{figure}

Prior work on  broadband AMM design remains constrained by predefined templates: existing heuristic optimizations~\cite{BendsoeSigmund2004TOBook,Molesky2018NatPhotonInverseDesign,jensen2011topology,AppliedAcoustics2023TOAMS,noguchi2021topology,jafar2018adaptive,johnson1997genetic,Li2012JASAGA} and data-driven regressors~\cite{Zhang2023ActaMechSinInverseAcoustic,Lv2024MaterialsAcousticMetasurfaceDL,Zhou2025SciRepBroadbandAbsDL,Zhu2025ActaAcusticaANNAbsorber,AppliedAcoustics2025CylPlateDL,AppliedAcoustics2026NewInvDL,JSV2025InvDL118789} first assume a structural template (e.g., a duct-cavity layout) and then tune its dimensions, so if a template cannot express the desired full-band response, parameter tuning cannot fix it. 
Meanwhile, although image-based generative models have shown potential in electromagnetic design for exploring structures beyond predefined templates~\cite{an2021multifunctional,zhao2025metagen}, their use in acoustic metamaterial design remains limited~\cite{Amirkulova2022ICA}.  A key  reason is that acoustic designs often require continuous fluid  pathways, whereas image-based models can generate broken paths, or detached pixels that violate such geometric constraints.  They also face a tradeoff between geometric precision and computational cost: high-resolution images are expensive to train on, while compressed  inputs lose fine structural details that are acoustically important. This leaves a clear  gap  for inverse-design frameworks that can explore diverse AMM structures while preserving the geometric precision and structural validity required by acoustics.

We present \sysname, a physics-guided, sequence-based  generative framework for AMM inverse design. Its core is a novel structured language that represents each AMM structure as a \textbf{sequence} of  acoustic primitives (e.g., \textit{Duct, Cavity}),   topological relations,  and geometric dimensions (e.g., \textit{Width, Length})  (Fig.~\ref{fig:fig1}). In this language, primitives form the vocabulary, and connection rules define the grammar. Compared with image-based representations, this language preserves geometry without relying on high-resolution pixel grids, reducing computational cost while retaining acoustically important details.
It also encodes structural connectivity through explicit relations between neighboring primitives, so generated structures maintain continuous fluid paths instead of broken  regions. Built on this representation, \sysname casts inverse design as a sequence-to-sequence generation task from target response  to structure sequence, without relying on templates. After supervised training on paired data, \sysname's generator is further refined with reinforcement learning, guided by a physics-based response solver, to encourage generating multiple valid structure sequences with low response error rather than overfitting to a single reference sequence in the training set.

\sssec{Language Representation of AMM Structures:} To move beyond fixed templates and image-based representations,  \sysname introduces a language for AMM structures that preserves the geometric precision and structural connectivity required by acoustics. This representation enables \sysname to cast inverse design as a sequence-to-sequence generation task. The key challenge is that an AMM structure is not merely a list of primitives, but a connected composition whose behavior depends on how those primitives are arranged. A naive  sequence representation that simply lists primitives (e.g., \texttt{[Duct}, \texttt{Neck}, \texttt{Cavity]}) is ambiguous: it does not specify whether the primitive is connected in series or as a side branch.
\sysname resolves this by introducing  relation tokens  (e.g., \texttt{[Shunt\_Start]},\texttt{[Shunt\_End]}) acting as delimiters, together with primitive tokens (e.g., \texttt{[Duct]}), so that connectivity is encoded explicitly in the sequence.  Each primitive token is further associated with numerical parameters that specify its geometric dimensions.  However, relation tokens alone are still insufficient, because AMM structures may contain  side branches,  and the ordering of tokens in the 1D language sequence must be explicitly defined.
\sysname therefore defines a depth-first grammar that determines the token order by writing out each side branch before continuing along the propagation direction. Together, the vocabulary, grammar, and numerical parameters provide a sequence representation of AMM structures.

\sssec{Balanced and High-Fidelity Dataset Construction:} With the language defined, \sysname constructs a large-scale dataset of  AMM sequences paired with their acoustic responses for  training. To ensure structural validity, we build a validity checker that enforces the language grammar and geometric constraints during data generation. A key goal in this dataset construction is broad topological coverage over  structures, but  randomly assembling primitives  biases the dataset toward simple structures, since more complex ones are more likely to be rejected by the validity checker due to spatial overlaps. To improve coverage, \sysname therefore adopts a complexity-balanced sampling strategy that maintains a uniform distribution over structural lengths (i.e., the total number of side-branches and main-path primitives).
Another challenge is  obtaining high-fidelity response  labels at scale. Finite element method (FEM) simulation (e.g., COMSOL) is accurate but expensive~\cite{ma2021deep} (3 s/sample), while analytical solvers like Transfer Matrix Method (TMM)~\cite{herrero2019matrix} are faster but less accurate, with an average mean squared error (MSE) of $0.116$ relative to COMSOL, especially for complex topologies with cross-sectional discontinuities. To resolve this, \sysname introduces an efficient FEM-intervened calibration mechanism that models these errors as geometry-dependent equivalent-length corrections. Calibrated on a small FEM pilot set, \sysname's  physical solver remains efficient (0.024 s/sample) while achieving high fidelity ($0.011$ average MSE), providing  reliable  labels.

\sssec{Physics-Guided Sequence Generation:} A central challenge in training a sequence-based generative model for AMM inverse design is the one-to-many nature of the task: multiple structures can realize the same target response, yet  supervised training tends to overfit to the reference sequence in the training set and penalize alternative valid designs. This limitation also matters in practice, because different valid structures may offer different trade-offs for downstream deployment, such as compactness or fabrication robustness. A natural solution is to augment supervised pretraining objective with a response-space loss by evaluating each generated candidate using \sysname's physical solver and validity checker, instead of relying solely on exact sequence matching. However, this involves non-differentiable validity checking, making end-to-end backpropagation  impractical.
To address this, \sysname adopts a two-stage training: supervised pretraining 
learns the structured AMM language, followed by reinforcement learning (RL)  to optimize sequence generation guided by the solver and the non-differentiable  checker. A key challenge in ensuring the effectiveness of this paradigm is preserving the coupling between discrete primitive tokens and continuous geometric parameters during generation, because independent prediction can produce semantically inconsistent token-parameter pairs. \sysname therefore introduces a dual-head architecture that conditions geometric parameter prediction on the generated primitive token. \sysname further combines structure-aware scalar reweighting with a validity-aware RL reward to improve response matching while reducing structural violations.

We implement \sysname as an end-to-end framework that maps target responses to valid structured sequences, from which AMM structures are deterministically reconstructed via cumulative geometric parameters~\cite{cai2013collision}. We validate the responses of \sysname's generated design using COMSOL simulations, and compare \sysname with five baselines,  including traditional optimization methods~\cite{zhu1997algorithm}, image-based generative models~\cite{zhao2025metagen}, across training efficiency, inference latency, validity, and response accuracy.



\sssec{Results: } Our results show that:
\begin{packeditemize}
	\item \sysname achieves a mean MSE of 0.0423 between the generated design’s response and the target response.
	
	\item Compared with image-based diffusion baselines and traditional optimization methods, \sysname improves accuracy by 60\% and 45\%, reduces
    inference latency by $4571\times$ and 662$\times$, respectively. \sysname 
    reduces training time by 7296$\times$ comparing with image-based diffusion.
\end{packeditemize}

\sssec{Contributions: }\sysname's core contributions include:

\begin{packeditemize}
	\item A  generative framework that reformulates acoustic metamaterial inverse design as a seq2seq translation task.
	
	\item A language for representing acoustic metamaterial structures for  inverse design. 
	
	\item A balanced, high-fidelity dataset enabled by an efficient calibrated physical solver.
	
	\item A generative model that addresses the one-to-many inverse design problem while promoting validity.
\end{packeditemize}


\section{Background}
\subsection{Acoustic Metamaterials}
Acoustic metamaterials (AMMs) are engineered sub-wavelength structures whose  geometries determine their transmission or reflection amplitude and phase across frequencies~\cite{fok2008acoustic,Assouar2018NRM}. An AMM unit is built from components such as labyrinthine ducts, cavities, and narrow necks; such unit can also be assembled into a larger acoustic metasurface~\cite{Ma2014NatMater,LiAssouar2016APL}.



\sssec{Design Pipeline.} Whether designing a standalone AMM or a larger metasurface, the pipeline is top-down: designers first convert a desired wave-control objective (e.g., high absorption, or focusing) into target responses for individual AMM structures,  specified as frequency-dependent phase and amplitude (e.g., transmission  coefficients). The inverse-design problem is then to map these target responses to geometries. Existing approaches often solve this problem by searching over predefined templates and tuning their parameters, with substantial expert input. However, for broadband targets, selecting a suitable template is itself difficult. \sysname instead  maps target responses directly to geometry design  without relying on 
predefined templates.

\subsection{Physical Solver for Acoustics}\label{sec:solver}
Numerical solvers such as FEM~\cite{ma2021deep} are accurate for AMM response computation but too expensive for data generation. 

\sssec{Analytical Physical Solver.} To accelerate computation, a common analytical alternative combines  equivalent circuit modeling (ECM)~\cite{bruneau2013fundamentals} and the transfer matrix method (TMM)~\cite{herrero2019matrix}. As shown in Fig.~\ref{fig:TMM}, ECM models localized resonant structures as impedance elements, while TMM models wave propagation through connected ducts by cascading transfer blocks $\mathbf{M}$. 
The analytical solver  decomposes an AMM into acoustic components (primitives) and models each by its dominant behavior. Localized resonant components, such as a narrow neck attached to a cavity, are represented as lumped impedance elements with ECM, while propagation-dominant units, such as straight ducts or coiling channels, are represented as transfer blocks with TMM. The response of the full structure is then computed by composing these component-level models according to the structure connectivity; details are given in Appendix~\ref{app:solver_details}. This abstraction is efficient for sub-wavelength AMM structures, but can lose accuracy at structural discontinuities such as duct steps or junctions.  \sysname addresses this limitation with an efficient FEM-intervened calibration, detailed in Sec.~\ref{sec:correction}.

\begin{table}[t]
	\renewcommand{\arraystretch}{0.9} 
	\centering
	\caption{Scale-wise performance of the diffusion model.}
	\vspace{-10pt}
	\scalebox{0.75}{
		\begin{tabular}{lcccc}
			\toprule[1.5pt]
			\textbf{Compression Scale} & \textbf{Original} & $\sim$\textbf{10x} & $\sim$\textbf{20x} & $\sim$\textbf{40x} \\
			\midrule
			Dimension (px) & $6848 \times 3200$  &$704 \times 336$ & $352 \times 176$ & $192 \times 96$ \\
			Ratio (1px) & $0.1\text{mm}$ &$1\text{mm}$ & $2\text{mm}$ & $4\text{mm}$ \\
			\begin{tabular}[c]{@{}l@{}}Collision Case \\ (out of 50k)\end{tabular} & - & 0 & 0 & 4254 \\
			\begin{tabular}[c]{@{}l@{}}Training Time \\ per Epoch (H100)\end{tabular} & N/A (OOM) & N/A (OOM) & $\sim$14hrs & $\sim$95mins \\
			MSE due to scaling & - & 0.0330 & 0.0564 & 0.0702 \\
			\bottomrule[1.5pt]
	\end{tabular}}
	\label{tab:scale_spec}
	\vspace{-20pt}
\end{table}
\begin{figure}[t]
	\centering

	\subfigure[Collision case when scale image]{
		\begin{minipage}{0.46\linewidth}
			\centering
			\includegraphics[width=0.7\textwidth]{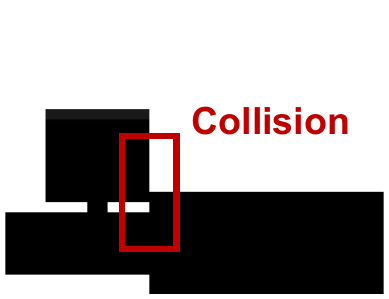}
		\end{minipage}
		\label{fig:collision}
	}
	\subfigure[Case of generated broken path]{
		\begin{minipage}{0.46\linewidth}
			\centering
			\includegraphics[width=0.7\textwidth]{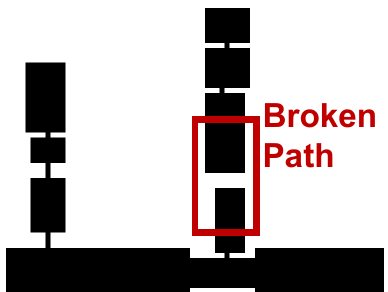}
			\vspace{2pt}
		\end{minipage}
		\label{fig:island}
	}
	\vspace{-15pt}
	\caption{Scaling images to reduce training overhead causes rounding-induced collisions, and diffusion can generate broken island structures.} 
	\label{fig:image}
	\vspace{-15pt}
\end{figure}

\subsection{Image-based Inverse Design} 
Image-based generative models have shown promise in electromagnetic design; for example, Metagen~\cite{zhao2025metagen} uses diffusion to explore structures beyond predefined templates. However, they are a poor fit for acoustic metamaterial inverse design, which requires both geometric precision and structural validity. High-resolution images  preserve fine geometry but incur prohibitive memory and training cost, including out-of-memory (OOM) errors on H100 GPUs in our study, while compressed images introduce  rounding artifacts that distort thin structures and creates collisions (Fig.~\ref{fig:collision}). Moreover, acoustic  structures require continuous fluid pathways, whereas Metagen~\cite{zhao2025metagen} produces broken paths (Fig.~\ref{fig:island}).  Our study, which adapts the Metagen~\cite{zhao2025metagen} backbone to  50,000 acoustic structures for the $100\text{--}1000\text{Hz}$ band, confirms this tradeoff; setup details are in Appendix~\ref{app:diffu} and results in Table~\ref{tab:scale_spec}. These limitations motivate a representation that explicitly preserves geometric dimensions and structural connectivity. 

\section{\sysname Overview}
\subsection{Technical Overview}
\begin{figure}[t]
	\centering
	\includegraphics[width=0.9\linewidth]{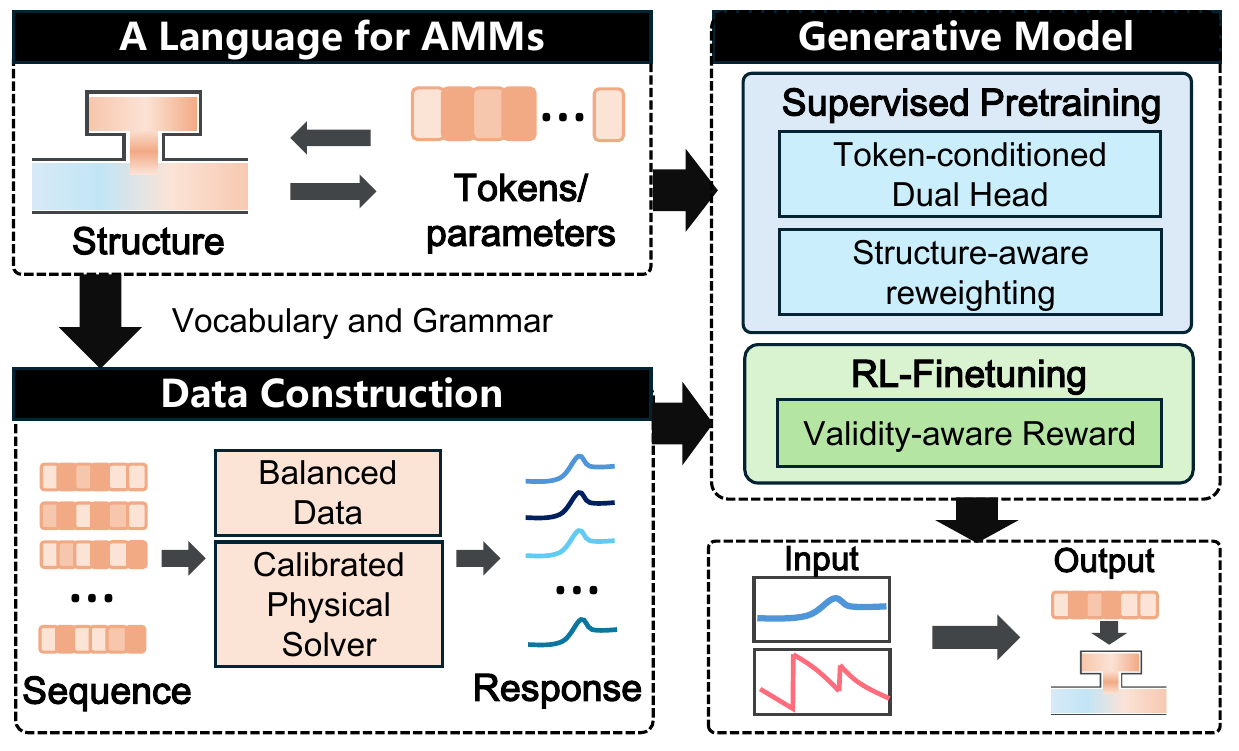}
	\vspace{-10pt}
	\caption{System overview of \sysname} 
	\label{fig:overview}
	\vspace{-15pt}
\end{figure}

\sysname introduces a structured sequence representation to formulate the AMM inverse design as a Seq2Seq translation problem: mapping a target spectral response, to a  sequence of acoustic primitives, their geometric dimensions, and their topological relationships. This approach leverages the compositional nature of AMM structures, where components such as ducts and neck-cavity branches combine hierarchically to determine the overall response. The sequence formulation also naturally mirrors wave propagation through connected acoustic primitives while preserving precise numerical parameters. By allowing the sequence length to vary with the target response, \sysname can accommodate diverse structural complexity beyond fixed templates. As shown in Fig.~\ref{fig:overview}, \sysname includes three coupled modules:

\sssec{(1) A Language for AMMs (Sec.~\ref{sec:language}).} AMMs couple discrete topology with geometry. Naive linear serialization loses hierarchy, leading to ambiguity. \sysname addresses this by designing a language representation that serializes designs as token-parameter pairs: primitive tokens represent structure semantics, and relation tokens encode  connectivity. 

\sssec{(2) High-fidelity Data Construction (Sec.~\ref{sec:dataset}).} Generating large-scale response-sequence training data with full-wave simulation is prohibitively expensive. \sysname instead combines an analytical physical solver with FEM-intervened correction and complexity-balanced sampling, enabling efficient data generation with calibrated fidelity.


\sssec{(3) Generative Model (Sec.~\ref{sec:generative}).} Acoustic inverse design is inherently one-to-many. However, supervised training alone tends to overfit to one reference sequence, while response matching is hard to optimize directly because operations like validity checking are non-differentiable. \sysname therefore adopts supervised pretraining followed by RL fine-tuning, allowing the model to first learn the structured design language and then explore alternative candidates under response-based feedback. Realizing this design requires addressing two additional challenges: primitive tokens and geometric parameters must be predicted jointly to avoid semantic mismatch, and generated structures must satisfy validity constraints such as dimensional bounds and spatial non-overlap. 
To this end, \sysname introduces a token-conditioned dual-head architecture, together with structure-aware reweighting and a validity-aware reward.


\subsection{Modeling Assumptions and Scope}
\sssec{Sub-wavelength Dimensions.} We assume cross-sectional dimensions of structure remain below the half-wavelength cutoff, ensuring that acoustic propagation is dominated by fundamental plane waves~\cite{bruneau2013fundamentals}. To maintain fidelity, \sysname accounts for boundary conditions and impedance transformations at geometric junctions (Sec.~\ref{sec:correction}).

\sssec{Acoustically Rigid Boundaries} For airborne acoustics, we assume high impedance contrast between commonly used materials (e.g., 3D-printed PLA or resin) and air ($Z_{\text{solid}} \gg Z_{\text{air}}$) allows us to treat walls as rigid reflectors, neglecting fluid-structure interaction.

\sssec{Operating Scope:} We focus on  linear, passive airborne acoustics, excluding nonlinear effects under high sound pressure levels.  We treat each AMM as an isolated structure and neglect coupling with the surrounding environment.

\section{A Language for AMMs}\label{sec:language}
This section  introduces a language representation for AMM structures that preserves the geometric precision and structural connectivity. Specifically, we define a vocabulary and a set of grammatical rules to represent an acoustic structure into a 1D sequence,  transforming the inverse design problem into a sequence generation task (showcase in Fig.~\ref{fig:language}).


\subsection{Vocabulary}
AMM structure is not merely a set of primitives, but a composition where the assembly logic defines the acoustic response. A simple linear arrangement can be ambiguous because it cannot indicate whether the primitives are connected in parallel or in series. To this end, our vocabulary $\mathcal{V}$ consists of three distinct categories, representing the building primitives, topological relationships, and geometric properties:

\sssec{Acoustic Primitive Tokens ($\mathcal{V}_{pri}$):} distinct acoustic primitives are represented by unique identifiers (e.g., \texttt{[Duct]}, \texttt{[Neck]}, \texttt{[Cavity]}, \texttt{[Coil]}). These primitives encompass the representation of most current airborne elements~\cite{fok2008acoustic}. 
	
\begin{figure}[t]
	\centering
	\includegraphics[width=0.8\linewidth]{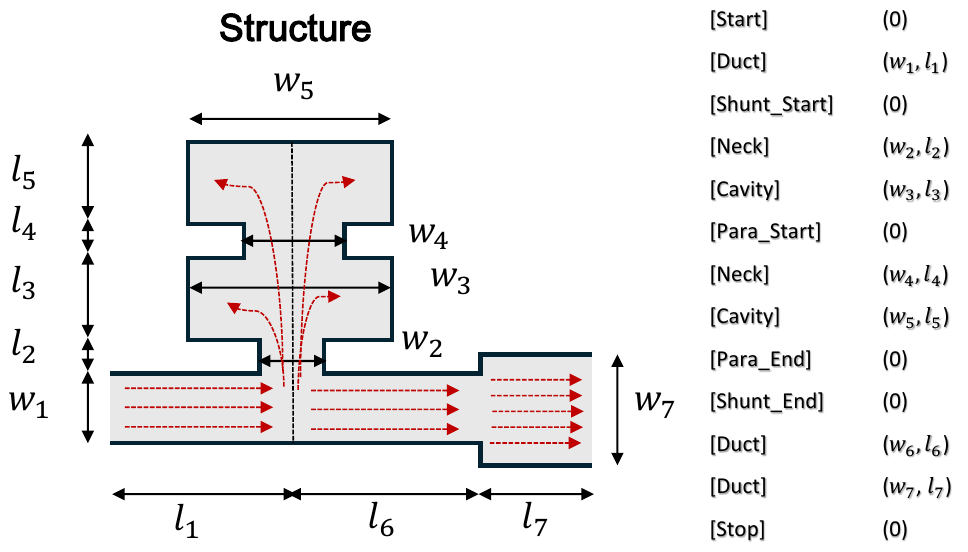}
	\vspace{-15pt}
	\caption{Example structural sequence in \sysname's language.} 
	\label{fig:language}
	\vspace{-20pt}
\end{figure}

\sssec{Relationship Tokens ($\mathcal{V}_{rel}$):} To encode topological connection and define sequence boundaries, we introduce a set of special delimiter tokens to connect primitive tokens:
\begin{itemize}[leftmargin=12pt,topsep=0pt]
	\item \texttt{[Start]} / \texttt{[Stop]}: Mark the global initiation and termination of the topological sequence, ensuring the generated design is complete and self-contained.
	
	\item \texttt{[Shunt\_Start]} / \texttt{[Shunt\_End]}: Delimit a subgraph connected to the main propagation path via a T-junction (parallel connection), effectively isolating the side-branch impedance from the main waveguide.
	
	\item \texttt{[Para\_Start]} / \texttt{[Para\_End]}: Encapsulate cascaded components within a side branch. Crucially, these tokens support \textbf{recursive nesting} (Fig.~\ref{fig:language}), allowing the language to represent multi-level acoustic networks or complex sub-structures within a single branch.
\end{itemize}
Notably, series connections are represented implicitly in tuple adjacency, effectively simplifying the sequence length.

\sssec{Numerical Parameters:} These encode continuous geometric dimensions. We represent the parameters of each primitive as $\mathbf{p} \in \mathbb{R}^n$, where $n$ depends on the primitive type: for example, $n=1$ for the radius of a 2D circle or 3D sphere, and $n\in{2,3}$ for 2D or 3D primitives with length, width, and height. This parameterization applies only to primitive tokens ($\mathcal{V}_{pri}$); relation tokens ($\mathcal{V}_{rel}$) are assigned a placeholder tuple $(0)$ for consistency. In the current implementation, each primitive is represented in 2D by width $w$ and length $l$, with $0.1\text{ mm}$ precision, and can later be extruded into a 3D structure for COMSOL simulation and fabrication.

\subsection{Grammar}\label{sec:grammar}
The syntax of \sysname governs how tokens from the vocabulary are assembled into valid sequences. To define the ordering of the tokens, we adopt a Depth-First Traversal (DFT) strategy~\cite{chen1991unified} to linearize the acoustic tree structure: Every time the traversal encounters a side-branch (e.g., a Helmholtz resonator shunt to a duct), the sequence enters the branch first (represented by \texttt{[Shunt\_Start]}) until the branch path is fully traversed (denoted by \texttt{[Shunt\_End]}), and then continues along the main path. In addition, \sysname also needs to obey the following grammars:

\sssec{Token-parameter Coupling:} To ensure semantic alignment, every token must be immediately followed by its corresponding parameters, ensuring that every acoustic primitive is fully defined by both its type and geometry.

\sssec{Structural Integrity:} The sequence must adhere to grammatical rules to define a valid structure.
\begin{itemize}[leftmargin=12pt,topsep=0pt]
	\item Balanced Delimiters: Relationship tokens must form balanced pairs, where every opening token (e.g., \texttt{[Shunt\_Start]}) must match a closing one (e.g., \texttt{[Shunt\_End]}).
	\item Non-empty Branches: A structural enclosure cannot be empty. To represent a physically meaningful impedance branch, there must be at least one valid component between any pair of start and end delimiters.
\end{itemize}

\sssec{Customization.} In addition to the base grammar, our framework empowers users to implement customized rules to refine the structural assembly. For instance, users can impose topological constraints on the resonator sequence, such as specifying the number of neck-cavity orders or mandating the exclusive use of Helmholtz resonators (neck-cavity pairs) while filtering out expansion chambers (i.e., cavity standalone). This modularity allows for an explicit and flexible delineation of the target design space based on specific acoustic requirements.

\section{Dataset Construction}\label{sec:dataset}
With the language defined, the next step for \sysname is to construct a large-scale dataset of valid AMMs sequences paired with responses for training. The current implementation focuses on transmissive acoustic design.

\subsection{Physical Validity Checker}\label{sec:checker}
To ensure that generated sequences correspond to valid structures including structural validity and non-overlap validity, we implement a validity checker including two phases:

\sssec{Structural Checker.} The structural checker outputs a binary validity score to evaluate sequences against two primary constraints: \textit{1) Grammar validity:} check each token to see if it comply with the grammar defined in Sec.~\ref{sec:grammar}; \textit{2) Geometry Bounds:} all parameters must fall within pre-defined ranges to maintain sub-wavelength assumptions.

\sssec{Overlap Checker.} The overlap checker verifies that the assembled geometry is spatially non-overlapping and dimensionally compatible. It first maps the sequence into a 2D layout, where each primitive is represented as an Axis-Aligned Bounding Box (AABB)~\cite{cai2013collision}. Main-path series blocks are placed along the lateral axis, while shunt blocks are anchored at connection points and extend outward. The checker then performs pairwise AABB intersection tests. The checker finally outputs a binary validity flag and an overlap metric defined as the total intersecting area. Details are provided in Appendix~\ref{app:checker}. 

\begin{figure}[t]
	\centering
	\subfigure[Structural Length.]{
		\centering
		\includegraphics[width=0.3\linewidth]{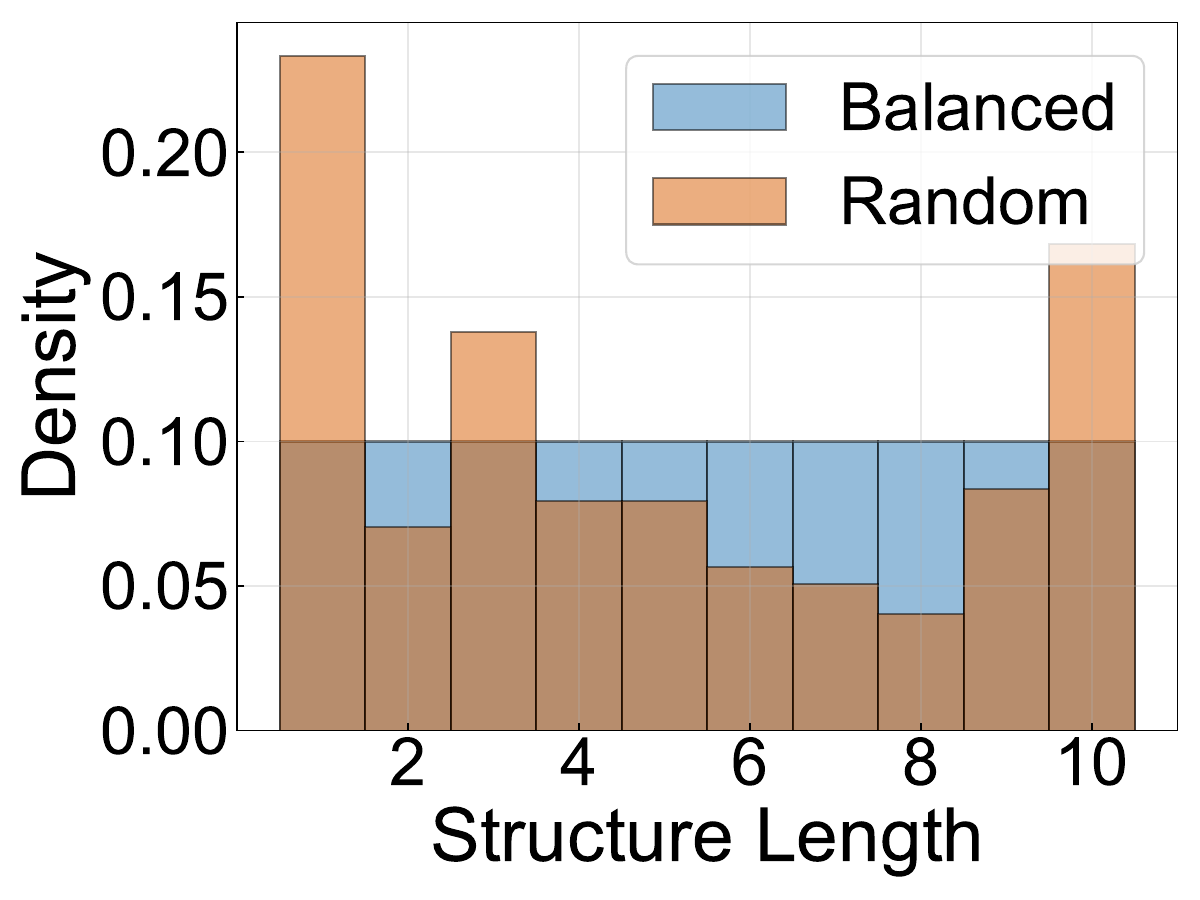}
		\label{fig:comp_dist}
	}\hspace{0pt}
	\subfigure[Sequence Length.]{
		\centering
		\includegraphics[width=0.3\linewidth]{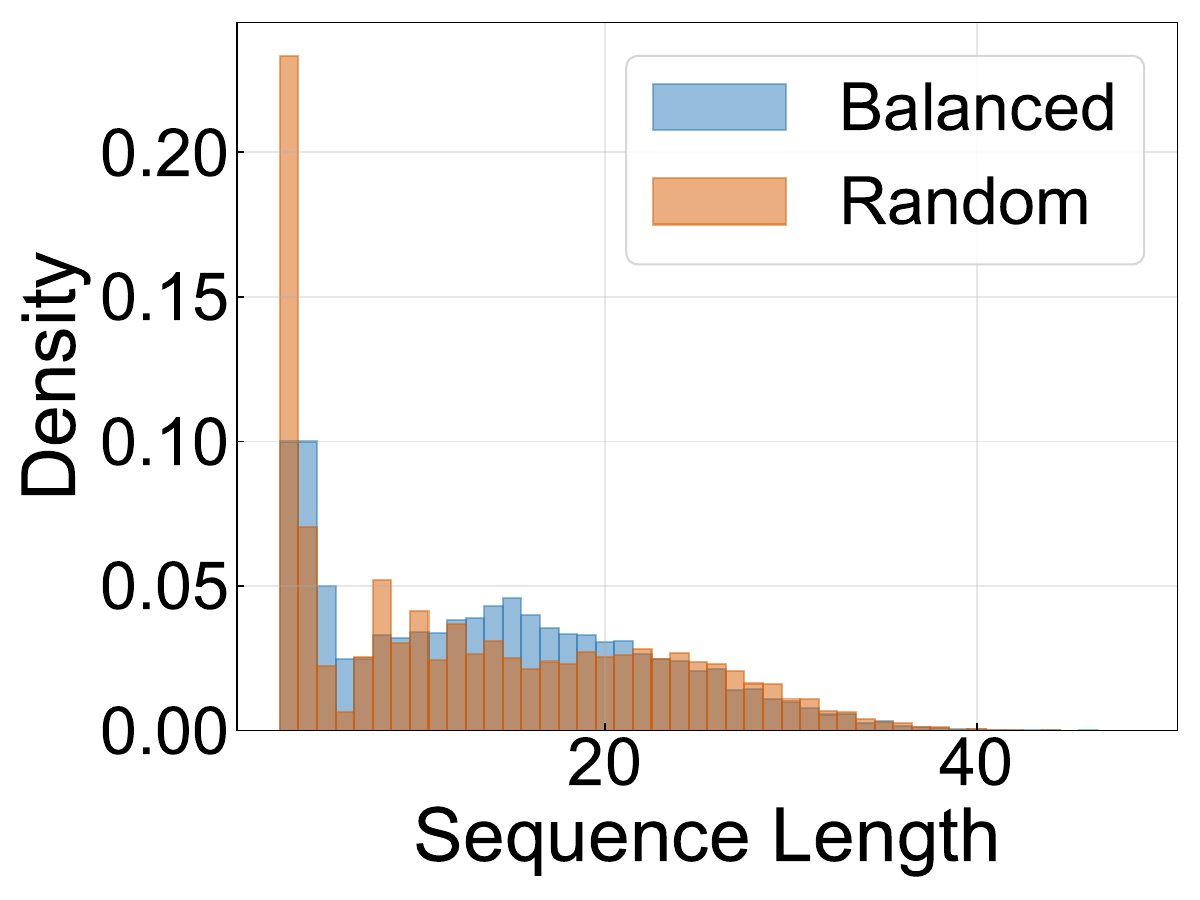}
		\label{fig:seq_dist}
	}\hspace{0pt}
	\subfigure[Relation.]{
		\centering
		\includegraphics[width=0.3\linewidth]{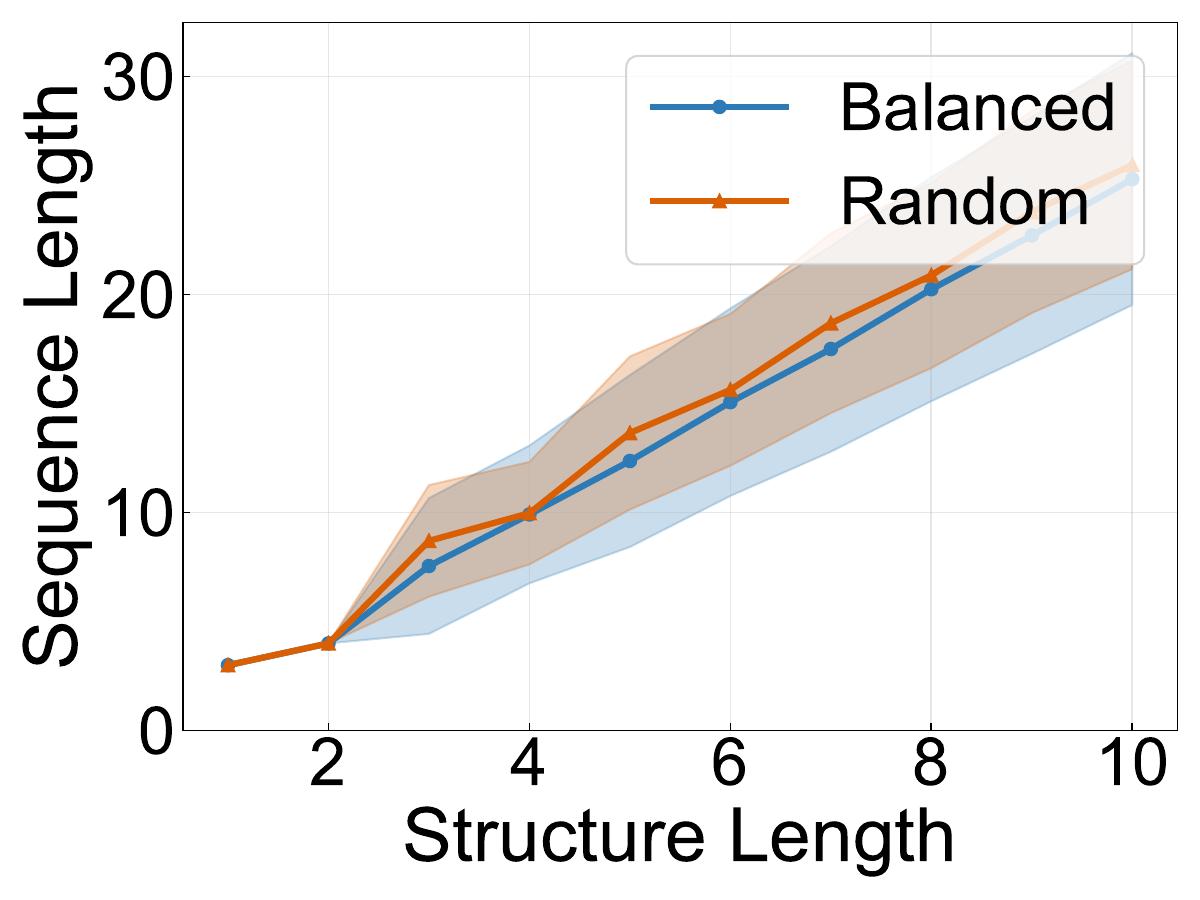}
		\label{fig:comp_vs_seq}
	}
	\vspace{-15pt}
	\caption{Distributions of component counts and sequence length in the generated dataset. } 
	\label{fig:hist}
	\vspace{-15pt}
\end{figure}

\subsection{Balanced Dataset Construction}\label{sec:baldataset}

A key goal in data construction is  broad topological coverage. However, random primitive assembly followed by validity checking biases the dataset toward short, simple structures, since longer sequences are more likely to be rejected due to spatial overlaps (Fig.~\ref{fig:seq_dist}). To mitigate this, we propose a complexity-balanced sampling strategy. 
We observe that sequence length scales approximately linearly with structural length, defined as the total number of side branches and main-path primitives (Fig.~\ref{fig:comp_vs_seq}). Thus, we use structural length as a proxy for topological complexity and balance the dataset accordingly.
Specifically, we discretize the structural lengths into  tiers and cap each tier at  $10\%$ of the dataset (Fig.~\ref{fig:comp_dist}). This improves coverage across complexity levels and reduces overfitting to frequent simple cases. Sec.~\ref{eva:balance} compares random and balanced datasets.


\subsection{Physical Solver Calibration}\label{sec:correction}
\sssec{Limits of the Existing Physical Solver.}
To generate large-scale training data efficiently, \sysname utilizes the  acoustic solver described in Sec.~\ref{sec:solver}. 
However, the solver relies on simplified 1D assumptions and becomes less accurate at \textit{abrupt geometric discontinuities}, such as sudden area expansions, T-junctions, and branch entrances, where evanescent modes and boundary-layer effects are neglected~\cite{herrero2019matrix}. As a result, directly using its outputs as labels introduces a mismatch with high-fidelity FEM simulations and can bias model training.

\sssec{Calibrating the Physical Solver with FEM.} Standard corrections often use empirical factors, such as adding an effective length $\delta l$ to a resonator neck to capture radiation effects, typically approximated as $\delta l \approx 0.6r-0.85r$~\cite{bruneau2013fundamentals},
where $r$ is the neck radius. However, such corrections assume simple geometries (e.g., circular cross-sections and single junction).  They do not transfer well to our setting, which includes arbitrarily nested branches and multiple interacting junctions, making static parameters ineffective. Using full FEM throughout would improve accuracy, but is too expensive for large-scale dataset generation.

To address this, \sysname develops a computationally efficient FEM-intervened calibration method that improves the fidelity of the physical solver while preserving analytical tractability. Specifically, we introduce a type-conditioned junction correction based on the local area ratio $\eta$ and the connection type $t$ (e.g., neck-cavity junction, T-junction, area-discontinuity, etc.). Instead of applying a global scaling factor, we augment the physical length of each component connected to a junction by a geometry-dependent term and define the effective acoustic length as $l_{\text{eff}} = l + \delta_{t}(\eta)\,w$,
where $l$ is the physical length, $w$ is a local transverse scale (i.e., width) used to convert original physical length into an equivalent length, and $\delta_t(\eta)$ is a learnable correction factor specific to type $t$. We calibrate $\delta_t(\eta)$ on a small pilot dataset $\mathcal{D}_{\text{calib}}$ by minimizing the spectral discrepancy between the calibrated analytical solver and the ground-truth FEM solver:
\begin{equation}
	\delta^\star = \arg\min_{\delta} \;
	\sum_{i \in \mathcal{D}_{\text{calib}}} \sum_{\omega \in \Omega}
	\left\| \mathbf{y}_{\text{ana}}^{(i)}\!\left(\mathbf{l} + \delta_{\mathbf{t}}\!\left(\boldsymbol{\eta}\right) \odot \mathbf{w}; \omega\right)
	- \mathbf{y}_{\text{FEM}}^{(i)}(\omega) \right\|_2^2 \nonumber
\end{equation}
where $\Omega$ denotes the set of frequencies, $\mathbf{y}(\omega)$ denotes the frequency response (including transmission magnitude and phase), $\odot$ is the element-wise product. We optimize this objective using the Nelder-Mead algorithm~\cite{singer2009nelder}.
To account for sampling stochasticity, we repeat calibration on five random subsets and keep the best configuration. The calibrated correction is then frozen and applied during full-scale generation, improving solver fidelity with negligible extra cost.


\sssec{Selection of Calibration Samples.} The same distribution skew discussed earlier also affects calibration: simple random sampling over-represents short sequences (Fig.~\ref{fig:seq_dist}) and can cause the learned correction $\delta_t^\star(\eta)$ to overfit to short structures. \sysname therefore adopts a length-stratified sampling, drawing uniformly across structural lengths in the balanced dataset (Fig.~\ref{fig:comp_dist}). Since  sequence length is approximately linear to the number of structural length (Fig.~\ref{fig:comp_vs_seq}), this yields a calibration set $\mathcal{D}_{\text{calib}}$ that better covers both short and long configurations, improving the robustness of the correction.

\sssec{Fidelity Analysis.} We evaluate solver fidelity on a hold-out FEM test set using the MSE in Eq.~\ref{eqn:mse}. A pilot set of 40 calibration samples is sufficient, as the MSE plateaus beyond this point, while calibration takes only about 300 seconds (Fig.~\ref{fig:MSE_time_FEM}). With this setting, \sysname achieves the best average MSE of $0.011$, compared to $0.014$ for random sampling, $0.040$ for empirical correction, and $0.116$ with no correction (Fig.~\ref{fig:FEM_compare}). Additional details are deferred to Appendix~\ref{app:fidelity}. This efficient FEM-intervened correction ensures that the physical solver retains both computational efficiency and high fidelity for rapid construction of a massive, topologically diverse dataset.

\section{Generative Model Design}\label{sec:generative}
\subsection{Model Overview}\label{sec:problem}
\sssec{Inverse Design Objective.} \sysname's model aims to map a target broadband amplitude and phase response to \sysname's  structured sequence  representation (Sec.~\ref{sec:language}).
Given a target response  sampled on the frequency grid $\Omega$,
$$
\mathbf{Y} = \{(T(\omega),\phi(\omega)):\omega\in\Omega\}
$$
the goal is to generate a variable-length sequence of structural-geometric tokens:  
$$
\mathbf{d}=(d_1,\ldots,d_L),\qquad d_t=(s_t,\mathbf{p}_t)
$$
where $s_t\in\mathcal{V}$ is a structural token in vocabulary and $\mathbf{p}_t\in\mathbb{R}^2$ is its   associated geometric parameter vector. The response of the generated sequence should yield a metamaterial whose response matches $\mathbf{Y}$ as closely as possible. 
The resulting complete sequence $\mathbf{d}$ is then be deterministically decoded into the final metamaterial layout via~\cite{cai2013collision}.

\begin{figure}[t]
	\centering
	\includegraphics[width=0.95\linewidth]{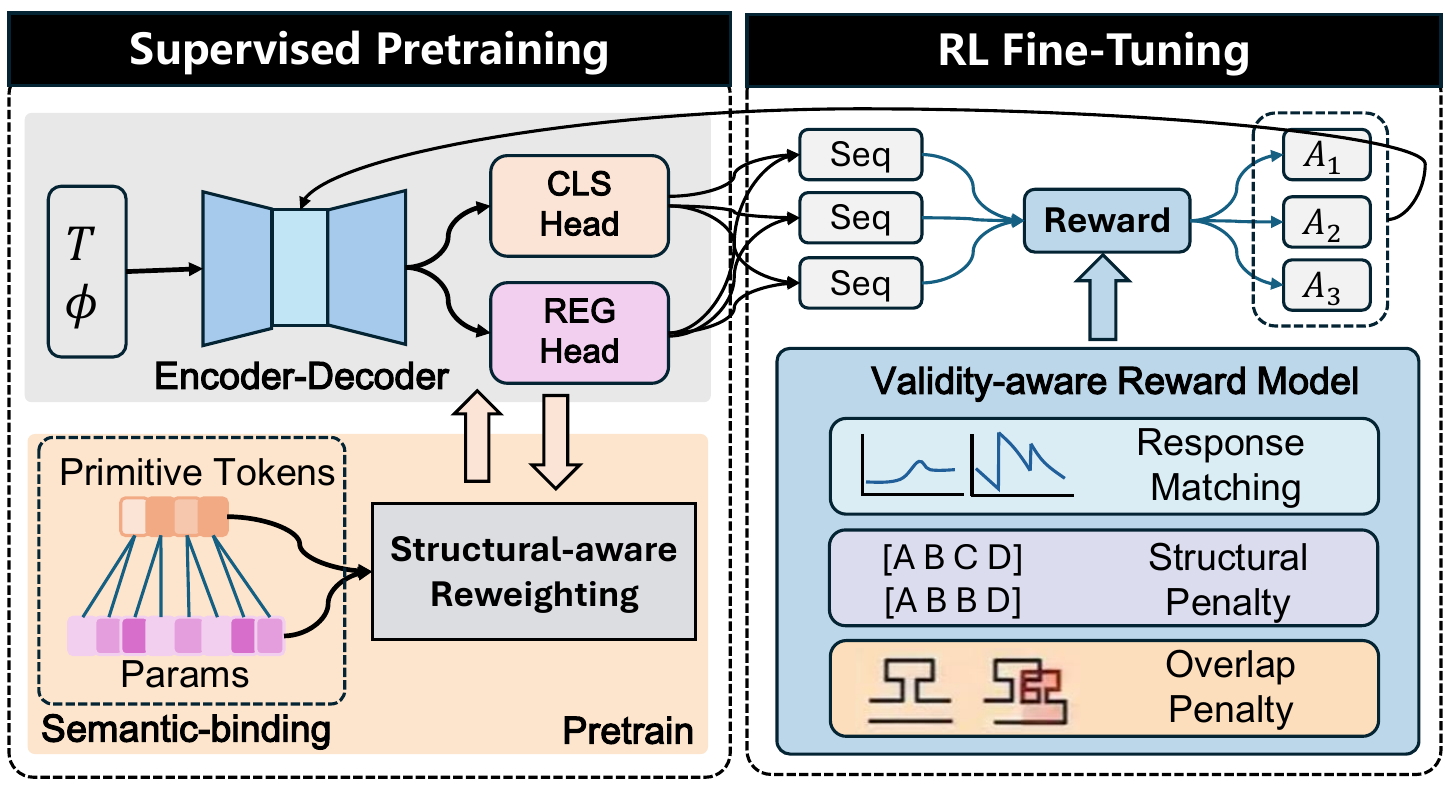}
	\vspace{-10pt}
	\caption{\sysname's Model Architecture} 
	\label{fig:model_structure}
	\vspace{-15pt}
\end{figure}

\sssec{Model Backbone.} \sysname uses a Transformer architecture~\cite{vaswani2017attention} as a conditional encoder-decoder backbone, as it can capture long-range dependencies in structured sequences.
The encoder condenses the target spectrum into a conditioning representation, while the decoder autoregressively generates the sequence via causal masking and cross-attention. However, several challenges remain: each step must jointly predict discrete structural tokens and continuous geometric parameters with semantic consistency, while local next-step prediction alone cannot ensure full-sequence validity, including grammar and geometric feasibility. Sec.~\ref{sec:pretrain}--\ref{sec:RL} present our solutions that address these challenges.



\sssec{Validity Requirements.} The sequences generated by \sysname must satisfy both structural and geometric validity constraints, as detailed in Sec.~\ref{sec:checker}. Detailed  formulation and handling are described in Sec.~\ref{sec:RL}.


\sssec{Training Paradigm.} The training of \sysname is inspired by large language models (LLMs), which also operate over structured sequence spaces with inherent one-to-many mappings~\cite{zhao2023survey}.
In particular, we tailor the training to acoustic metamaterial inverse design. Supervised pretraining (Sec.~\ref{sec:pretrain}) provides a strong initialization by enabling the model to learn topological syntax and token-parameter coupling from data. RL fine-tuning (Sec.~\ref{sec:RL}) then shifts the objective from imitating training sequences to directly improving target-response matching, while encouraging exploration of multiple valid candidates beyond the deterministic bias of the training data. 

\subsection{Initialization: Supervised Pretraining}\label{sec:pretrain}
Supervised pretraining can provide stable initialization for fine-tuning~\cite{schmied2023learning}. However, standard autoregressive Transformers~\cite{vaswani2017attention} are insufficient for \sysname: they struggle to maintain semantic consistency across joint discrete-continuous predictions, and token-wise likelihood fails to capture global grammatical quality.





\sssec{Token-Parameter Semantic Binding.}  A conventional NLP approach~\cite{spithourakis2018numeracy} would discretize continuous geometric parameters into symbols, which causes combinatorial explosion and discards ordinal structure. A naive dual-head design, where token classification and parameter regression are decoupled, can also produce semantic mismatch; for example, the model may predict a neck token but regress cavity-sized dimensions. To address this, we introduce a token-conditioned dual-head structure that makes geometric prediction explicitly conditioned on the selected structural token. At step $t$ with hidden state $h_t$, the parameter head projects $h_t$ into a comprehensive latent matrix $\mathbf{Z}_t \in \mathbb{R}^{|\mathcal{V}_{pri}| \times 2}$, where $\mathcal{V}_{pri}$ is the primitive token in vocabulary. In our implementation, each primitive uses two parameters, width and length. Each row in $\mathbf{z}_t^{(s)}$  represents the  dimensions  hypothesized exclusively for token $s$. Once the classification head selects the primitive token $s_{pri} \in \mathcal{V}_{pri}$, the final geometric latent $\mathbf{z}_t$ is deterministically gathered by indexing the matrix:
$\mathbf{z}_t=\mathbf{Z}_t[s_t]$. 
This  binds structure and geometry by construction, ensuring that the predicted dimensions are conditioned on the selected token.

\sssec{Structure-aware Scalar Reweighting.}  Local next-token likelihood in Transformers does not enforce global structural validity; a sequence can violate grammar constraints.  To address this limitation, we introduce a novel structure-aware scalar reweighting mechanism that dynamically penalizes the model based on the overall quality of the full-sequence. 
Specifically, for each training sample, we let the model generate a full sequence $\mathbf{s}^{\mathrm{pred}}$, and then measure its structural quality using two criteria.


\begin{itemize}[leftmargin=*, itemindent=0em, topsep=0pt, partopsep=0pt, parsep=0pt, itemsep=0pt]
\item \textit{Structural Mismatch.} We compute the token-level Levenshtein distance~\cite{levenshtein1966binary}  between the predicted and reference sequences $\mathbf{s}^{\mathrm{gt}}$, denoted by $d_{\mathrm{edit}}\!\left(\mathbf{s}^{\mathrm{pred}}, \mathbf{s}^{\mathrm{gt}}\right)$. Unlike standard cross-entropy, which compares tokens at the exact same index and over-penalizes  shifted alignments, this edit distance captures holistic structural similarity under insertions, deletions, and substitutions.


\item \textit{Number of Grammar Violations.} We use the checker in Sec.~\ref{sec:checker} to count grammar violations $N_{\mathrm{viol}}(\mathbf{s}^{\mathrm{pred}})$. This complements edit distance because a sequence may differ from the reference yet still be 
grammatically valid.

\end{itemize}
	
We combine these two terms into a dynamic scalar weight $w_{\mathrm{struct}}$, normalized by the predicted sequence length $L^{\mathrm{pred}}$:
\begin{equation}
w_{\mathrm{struct}}
=
\frac{
	d_{\mathrm{edit}}\!\left(\mathbf{s}^{\mathrm{pred}}, \mathbf{s}^{\mathrm{gt}}\right)
	+
	\alpha\,
	N_{\mathrm{viol}}\!\left(\mathbf{s}^{\mathrm{pred}}\right)
}{
	L^{\mathrm{pred}}
} \nonumber
\label{eq:structure_weight}
\end{equation}
where $\alpha$ is empirically set to 20 for controlling the relative penalty of grammar invalidity. When the generated sequence is highly divergent or grammatically invalid, $w_{\mathrm{struct}}$ increases, forcing stronger learning from the reference sequence. When the sequence remains structurally reasonable, the penalty is smaller. This encourages structurally valid and reference-aligned generation during pretraining, yielding a better initialization for subsequent RL fine-tuning.


\sssec{Geometry Boundary.} Unconstrained regression can  produce dimensions that violate sub-wavelength assumptions or fabrication limits. We therefore apply a token-specific Sigmoid mapping~\cite{cybenko1989approximation} to the unbounded latent parameters,
scaling them into predefined physically admissible ranges for each selected primitive token $s_{pri}$.
This enforces valid parameter bounds during generation while improving optimization stability in a normalized latent space.

\sssec{Putting It Together: Pretraining Loss.} The supervised objective combines cross-entropy loss for token classification ($\mathcal{L}_{cls}$) and L1 Loss for parameter regression ($\mathcal{L}_{reg}$). The reweighted supervised objective is: 
\begin{equation}
\mathcal{L}_{\mathrm{sup}}^{\mathrm{scaled}}
=
w_{\mathrm{struct}}\,\mathcal{L}_{\mathrm{cls}}
+
\lambda \mathcal{L}_{\mathrm{reg}}. 
\label{eq:scaled_supervised_loss}
\end{equation}
and
\begin{equation}
\mathcal{L}_{\mathrm{cls}}
=
\sum_{t=1}^{L}
\mathrm{CE}\!\left(
P(s_t \mid d_{<t}, \mathbf{Y}),
s_t^{\mathrm{gt}}
\right), \mathcal{L}_{\mathrm{reg}}
=
\sum_{t \in \mathcal{I}_{\mathrm{pri}}}
\left\|
\hat{\mathbf{p}}_t - \mathbf{p}_t^{\mathrm{gt}}
\right\|_1 \nonumber
\end{equation}
where $\mathcal{I}_{\mathrm{pri}}$ indexes time steps corresponding to primitive token (i.e., tokens that carry parameters), and other tokens do not contribute to regression. $\lambda$ balances the two terms. Minimizing Eqn.~\ref{eq:scaled_supervised_loss} provides a stable initialization for the subsequent RL fine-tuning stage. This objective still cannot guarantee spatial non-overlap, since overlap depends on the joint configuration of all parameters in the sequence. We address this in RL fine-tuning (Sec.~\ref{sec:RL}).

\subsection{RL Fine-Tuning with Physics Guidance}\label{sec:RL}

While supervised pretraining provides a stable initialization by teaching the model the structured language, token-parameter alignment, and basic grammatical validity, it still mainly learns to imitate a single reference structure from the training data. This is limiting because acoustic inverse design is inherently one-to-many: multiple valid structures can realize similar target responses while differing in topology or geometry. This flexibility can also be beneficial, as different valid designs may offer different trade-offs for downstream use, such as fabrication robustness or material usage.


Top-$k$ sampling~\cite{li2020sampling} can generate multiple candidates  by stochastically selecting from the most probable token subspace at each step. However, when used with a supervised model in Sec.~\ref{sec:pretrain}, it still inherits the bias of reference-sequence training.  A more appropriate objective is to score candidates by their realized acoustic responses rather than by reference-sequence matching. In principle, one could augment supervised training with a response-space physical loss, but direct backpropagation is impractical due to non-differentiable validity checking and constraint evaluation.

\sssec{Objective Shift: From Sequence Matching to Response Matching.}  \sysname addresses this by fine-tuning the pretrained model with RL.
Unlike top-k sampling from the pretrained model, \sysname optimizes a response-oriented objective (Eqn.~\ref{eq:rl_response_loss}) that favors candidates with lower  MSE, rather than better sequence matching. This enables exploration of new designs that differ from the training reference yet realize similar target responses.  Moreover, RL does not require gradients through the full pipeline: it optimizes a scalar reward that combines response matching with validity penalties, discouraging physically invalid samples.


The pretrained model now is treated as a conditional policy. Given a target response $\mathbf{Y}$ and  a generated design $\mathbf{d}$, the calibrated physical solver $\mathcal{F}$ computes the corresponding transmission $T$ and phase $\phi$ , denoted by  $\hat{\mathbf{Y}}=\mathcal{F}(\mathbf{d})$. We measure discrepancy by using MSE for transmission and circular MSE for phase:
\begin{equation}
	\mathcal{L}_{\mathrm{phy}}(\hat{\mathbf{Y}},\mathbf{Y})
	=
	\frac{1}{|\Omega|} \sum_{\omega\in\Omega}
	\left(
	\bigl(\hat{T}(\omega)-T(\omega)\bigr)^2
	+
	\frac{1}{\pi^2}
	\Delta\phi(\omega)^2\right)
	\label{eq:rl_response_loss}
\end{equation}
where $\Delta\phi(\omega)
=
\mathrm{wrap}\!\bigl(\hat{\phi}(\omega)-\phi(\omega)\bigr)$. 
The term $\frac{1}{\pi^2}$ scales the phase term comparable to the transmission term because they are equally important. This response loss lets \sysname favor alternative designs when they achieve better amplitude-phase matching than the reference sequence.

\sssec{Validity-aware Reward Model.} Beyond response matching, \sysname must generate sequences that satisfy physical validity constraints, including structural grammar and spatial non-overlap (Sec.~\ref{sec:checker}). 
To guide RL toward valid designs, we use the checker in Sec.~\ref{sec:checker} to detect overlap and measure its extent. Because a binary flag cannot capture overlap severity, we further use a continuous hinge penalty ($\mathcal{L}_{\mathrm{hinge}} (\cdot)$)~\cite{gentile1998linear} to quantify the magnitude of overlap. 
 We also use a binary indicator $\mathbb{I}{\mathrm{g}}$ to penalize any sequence that violates the grammar rules in Sec.~\ref{sec:checker}.
Combining response matching with these validity constraints, we define the RL reward as:
\begin{equation}
	R(\mathbf{d},\mathbf{Y})
	=
	-\,\mathcal{L}_{\mathrm{phy}}\!\bigl(\mathcal{F}(\mathbf{d}),\mathbf{Y}\bigr)
	- \lambda_{\mathrm{g}} \mathbb{I}_{\mathrm{g}}(\mathbf{d})
	- \lambda_{\mathrm{ov}} \mathcal{L}_{\mathrm{hinge}} (\mathbf{d}) 
	\label{eq:rl_reward}
\end{equation}
where $\lambda{\mathrm{g}}$ and $\lambda_{\mathrm{ov}}$ control the relative penalty weights (evaluated in Sec.~\ref{subsec:params}). This reward discourages invalid designs while guiding RL toward accurate and feasible structures.

\sssec{Physical Solver Backend.} The reward evaluator $\mathcal{F}$ can be instantiated either as a fast, data-driven forward surrogate model (Appendix.~\ref{app:forw}), which serves as a proxy for the physical solver, or as the calibrated physical solver in Sec.~\ref{sec:correction}. Both backends integrate naturally into the same response-space reward formulation.

\sssec{Group-based Exploration and RL Policy Update.}  We adopt GRPO~\cite{shao2024deepseekmath} to stabilize RL over coupled token-parameter outputs. For each target, the model samples a group of candidate sequences, evaluates them jointly, and updates the policy using group-relative rewards. This is well suited to inverse design, where multiple valid structures may realize the same target response. Details are described in Appendix~\ref{app:grpo}.


\sssec{Parameter Update.} During the RL, we update the full parameters of the pretrained model, as validated in Sec.~\ref{subsec:params}. 

\sssec{Inference:} At inference time, \sysname reflects the one-to-many nature of inverse design by generating multiple candidate sequences for the same target response $\mathbf{Y}$ through top-k token sampling. 
This is more effective after RL fine-tuning, as the objective shifts from reference-sequence matching to response-oriented optimization, encouraging structurally different candidates that still match the target response well. In our current evaluation, we select the one with the lowest response error.  Finally, the selected sequence is deterministically decoded into 2D spatial coordinates to form the final acoustic metamaterial layout.

\section{Implementation}\label{sec:impl}
\begin{table}[t]
	\renewcommand{\arraystretch}{1}
	\centering
	\caption{Dataset distribution.}
	\vspace{-10pt}
	\scalebox{0.7}{
		\begin{tabular}{l|l|c|c}
			\toprule[2pt]
			\multicolumn{2}{c|}{\textbf{Frequency Range}} & \textbf{100Hz--1kHz} & \textbf{16kHz--24kHz} \\
			\midrule[1.2pt]
			\multirow{4}{*}{\makecell[l]{Acoustic\\Primitive\\(size:mm)}}
			& \texttt{[Duct]}
			& $l \in [20,80],\ w \in [20,40]$
			& $l \in [4,12],\ w \in [2,6]$ \\
			\cline{2-4}
			
			& \texttt{[Cavity]}
			& $l \in [20,100],\ w \in [10,50]$
			& $l \in [0.5,10],\ w \in [1,4]$ \\
			\cline{2-4}
			
			& \texttt{[Neck]}
			& $l \in [2,15],\ w \in [2,8]$
			& $l \in [0.3,1.0],\ w \in [0.5,1.0]$ \\
			\cline{2-4}
			
			& \texttt{[Coil]}
			& $l \in [50,150],\ w \in [20,40]$
			& $l \in [5,20],\ w \in [2,6]$ \\
			\hline
			
			\multicolumn{2}{l|}{Structural length} & $1 \sim 10$ & $1 \sim 10$ \\
			\hline
			\multicolumn{2}{l|}{Helmholtz max order} & 4 & 2 \\
			\hline
			\multicolumn{2}{l|}{Num.\ of frequency bins} & 100 & 100 \\
			\hline
			\multicolumn{2}{l|}{Resolution (mm)} & 0.1 & 0.1 \\
			\hline
			\multicolumn{2}{l|}{Constraints} & \multicolumn{2}{c}{If \texttt{[Coil]} exists, it should be the last primitive.} \\
			\bottomrule[2pt]
	\end{tabular}}
	\label{tab:data_spec}
	\vspace{-20pt}
\end{table}

\sssec{Dataset Generation and Distribution:} We construct  two datasets using calibrated analytical physical solver to cover two acoustic regimes: (i) a low-frequency band (100 Hz–1 kHz) typically for noise absorption; and (ii) a high-frequency band (16–24 kHz) commonly for ultrasonic sensing and communication.  Each dataset comprises 50,000 balanced samples with structural lengths  from 1 to 10. The generation runs on a standard workstation (Intel i7-12700 CPU, 16 GB RAM), and completes each dataset in under 20 minutes, providing a $126\times$ speedup over FEM simulators (about 42 hours in COMSOL). The data statistics are summarized in Table.~\ref{tab:data_spec}. 


\sssec{Train-test split:} We randomly split the dataset into training, validation, and test sets with a ratio of 7:1.5:1.5. The test set contains only sequences unseen during training to prevent data leakage. This fixed split is used for all baselines.

\sssec{Model Implementation:} Our model uses a Transformer  with a 256-dimensional latent space. The target response is encoded by an MLP~\cite{popescu2009multilayer} with GELU activation~\cite{lee2023gelu} into a latent embedding. This is then processed by a 4-layer Transformer decoder following  BERT-Mini~\cite{devlin2019bert} to generate the structural sequence. During pretraining, we use Adam with a learning rate of $1.5\times10^{-3}$ and batch size 1024 for up to 200 epochs with early stopping patience of 30 epochs. During RL, we use GRPO with Adam, learning rate $8\times10^{-5}$, batch size 256 and group size 4 for up to 200 epochs with early stopping patience of 20 epochs. We implement the model  in Python 3.10.19 and PyTorch 2.5.1 with CUDA 12.1. All experiments run on one NVIDIA H100 GPU with 96 GB memory.

\sssec{From Sequence to 3D Structure:} The generated sequence is mapped to 2D layouts via Axis-Aligned Bounding Boxes (AABB)~\cite{cai2013collision}. This 2D acoustic path is then converted into a 3D structure by assigning it a height that remains below the sub-wavelength scale. To obtain the manufacturable solid geometry, walls are extruded along the boundaries of the airborne path while preserving the internal air channel. The resulting geometry can be directly imported into COMSOL for FEM simulation and exported for fabrication.




\section{Evaluation}

\begin{figure}[t]
	\centering
	\subfigure[MSE.]{
		\centering
		\includegraphics[width=0.45\linewidth]{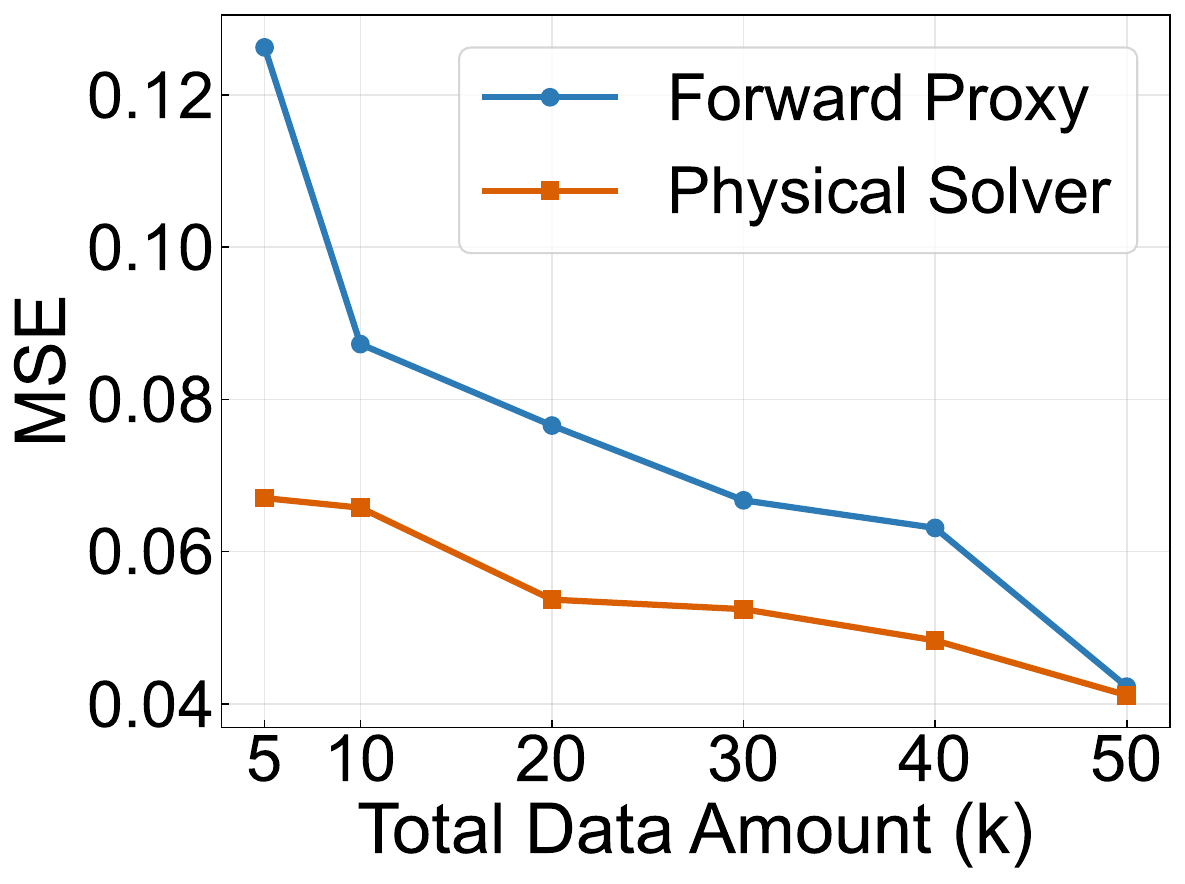}
		\label{fig:MSE_dataset}
	}\hfill
	\subfigure[Training time per epoch.]{
		\centering
		\includegraphics[width=0.45\linewidth]{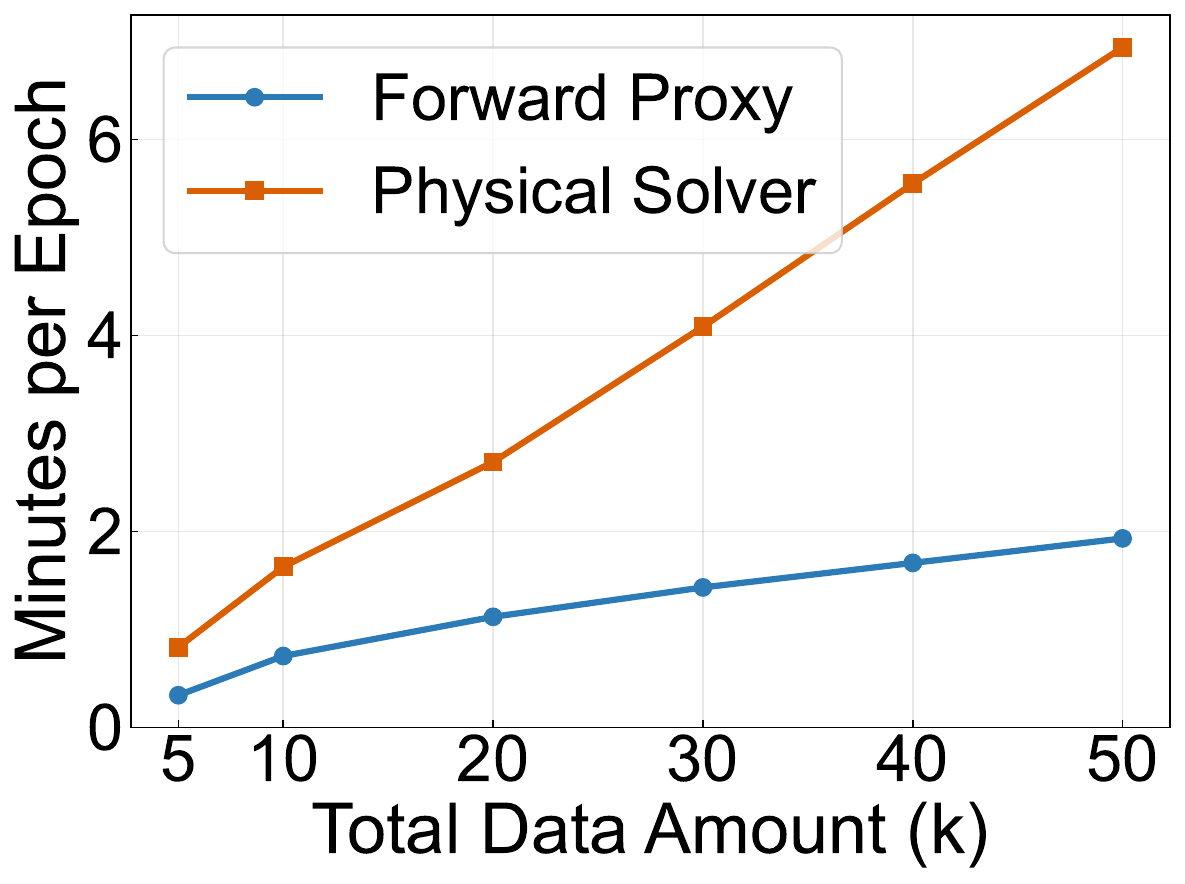}
		\label{fig:time_dataset}
	}
	\vspace{-15pt}
	\caption{Performance against the training set volume.} 
	\label{fig:volume}
	\vspace{-15pt}
\end{figure}

\subsection{Evaluation Methodology}\label{sec:eva_method}
\sssec{Evaluation Metrics:} We evaluate \sysname on two datasets (high-frequency and low-frequency), each containing 7,500 test samples. The primary metric is the average Mean Squared Error (MSE) against ground-truth FEM responses, calculated as the sum of transmission MSE and circular phase MSEs.
\begin{equation}
\small
\mathcal{E}_{\text{MSE}} \;=\; \frac{1}{|\Omega|} \sum_{\omega\in\Omega}
\left(
\bigl(\hat{T}(\omega)-T(\omega)\bigr)^2
+
\frac{1}{\pi^2}
\Delta\phi(\omega)^2\right)
\label{eqn:mse}
\end{equation}
where $\Delta\phi(\omega)
=
\mathrm{wrap} \!\bigl(\hat{\phi}(\omega)-\phi(\omega)\bigr)$. We also report the training time, inference time, and invalid count in test set.

\sssec{Baselines:} We compare \sysname against five baselines: 1) standard supervised learning on the training dataset without RL; 2) an image-based diffusion model with 40x compression; 3) a tandem network that augments the supervised model with a response-space loss; 4) RL from scratch without supervised learning; 5) a fixed-template baseline that simulates a traditional inverse-design workflow: 
samples are first categorized by the number of spectral peaks in their amplitude responses. For each test target, we then randomly select three training-set structures from the same category as \textit{templates} and optimize their geometric parameters using~\cite{zhu1997algorithm}. We report the best result among the three candidates.


\subsection{Microbenchmark}\label{sec:micro}
\sssec{Hyperparameters.}\label{subsec:params}
We study the impact of key hyperparameters in \sysname's training pipeline using validation-set MSE, including the RL fine-tuning scope, the weights of the structural invalidity and overlap penalties, top-$k$ sampling, and parameter perturbation scale. The study shows that full-model fine-tuning performs best, the two penalty terms should be weighted equally, top-$k$ sampling with ($k=2$) is sufficient, and moderate parameter exploration works best ($\sigma_p=0.5$). We use these settings in all subsequent experiments. Full results are provided in Appendix~\ref{app:hype}.

\sssec{Training Set Volume.} Model performance depends on training data volume, but larger datasets also increase computational cost. \sysname supports different physical evaluator backends. We compare a fast surrogate model trained on the training set (App.~\ref{app:forw}) with a precise but relatively slower calibrated physical solver. We vary the dataset size from $5\text{k}$ to $50\text{k}$ samples, using 70\% for training. All models are evaluated on the same 7500 samples. We report the MSE and the training time (including forward model training) per epoch. 

Fig.~\ref{fig:volume} shows that both mechanisms improve as the data scale increases. At smaller scales ($5k$-$40k$), the physical solver outperforms the forward model because the forward model remains data-limited in  learning the sequence-to-response mapping. However, the physical solver is slower, as it relies on sequential CPU execution,  whereas the forward surrogate runs on GPU. We note that  at  $50k$ samples, the forward model matches the physical solver's accuracy, while achieving  about $3\times$ higher efficiency. Therefore, we use the forward evaluator in subsequent experiments. 

\sssec{Balanced versus Random Dataset.}\label{eva:balance} We compare our balancing strategy (Sec.~\ref{sec:dataset}) against random generation. Balancing yields lower MSE, with performance gains scaling with structural complexity. Details are described in Appendix~\ref{app:ban_dataset}.

\begin{figure}[t]
	\centering
	\subfigure[CDF of MSE.]{
		\centering
		\includegraphics[width=0.465\linewidth]{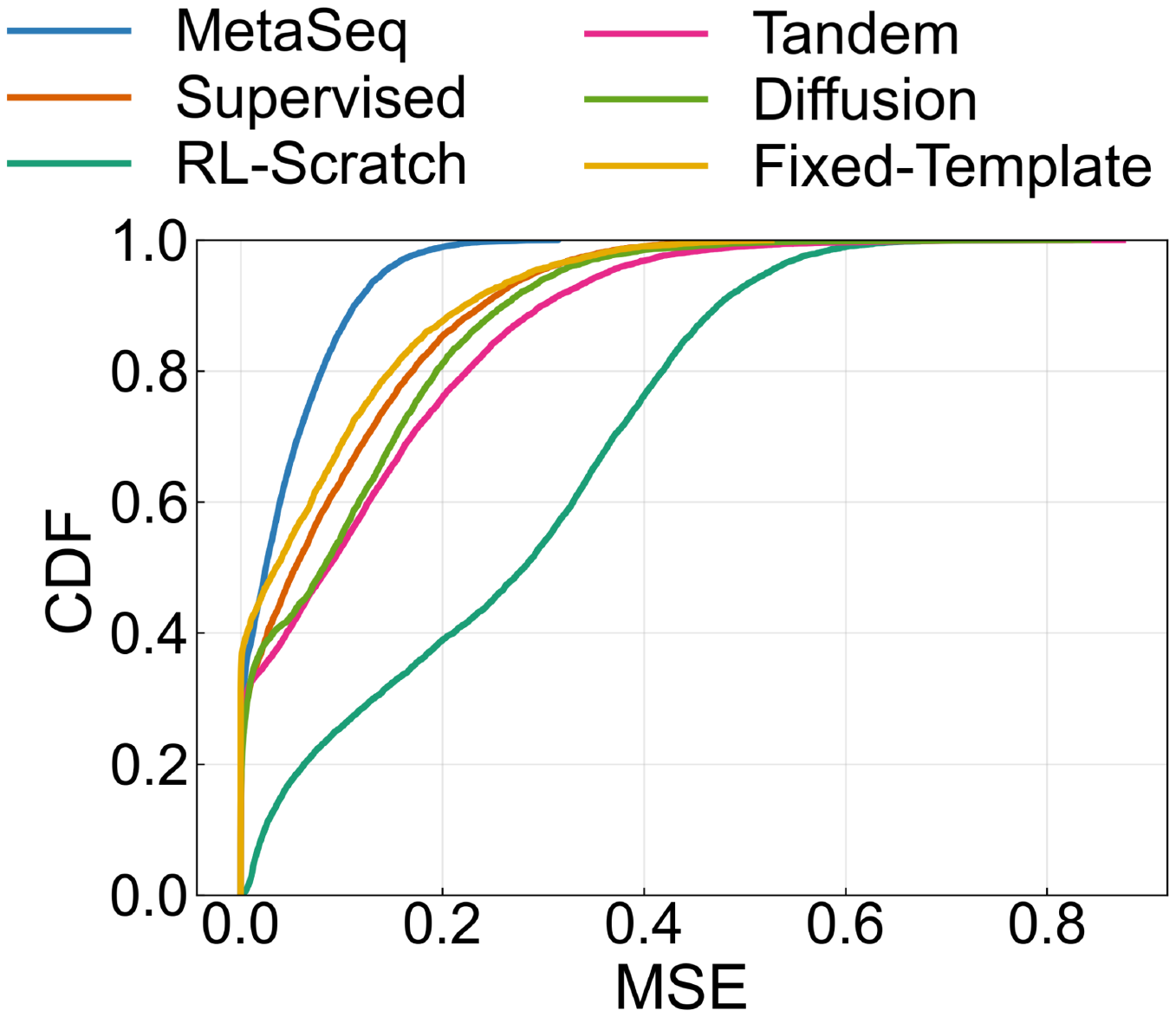}
		\label{fig:CDF_overall}
	}\hspace{-10pt}
	\subfigure[MSE against structural length.]{
		\centering
		\includegraphics[width=0.465\linewidth]{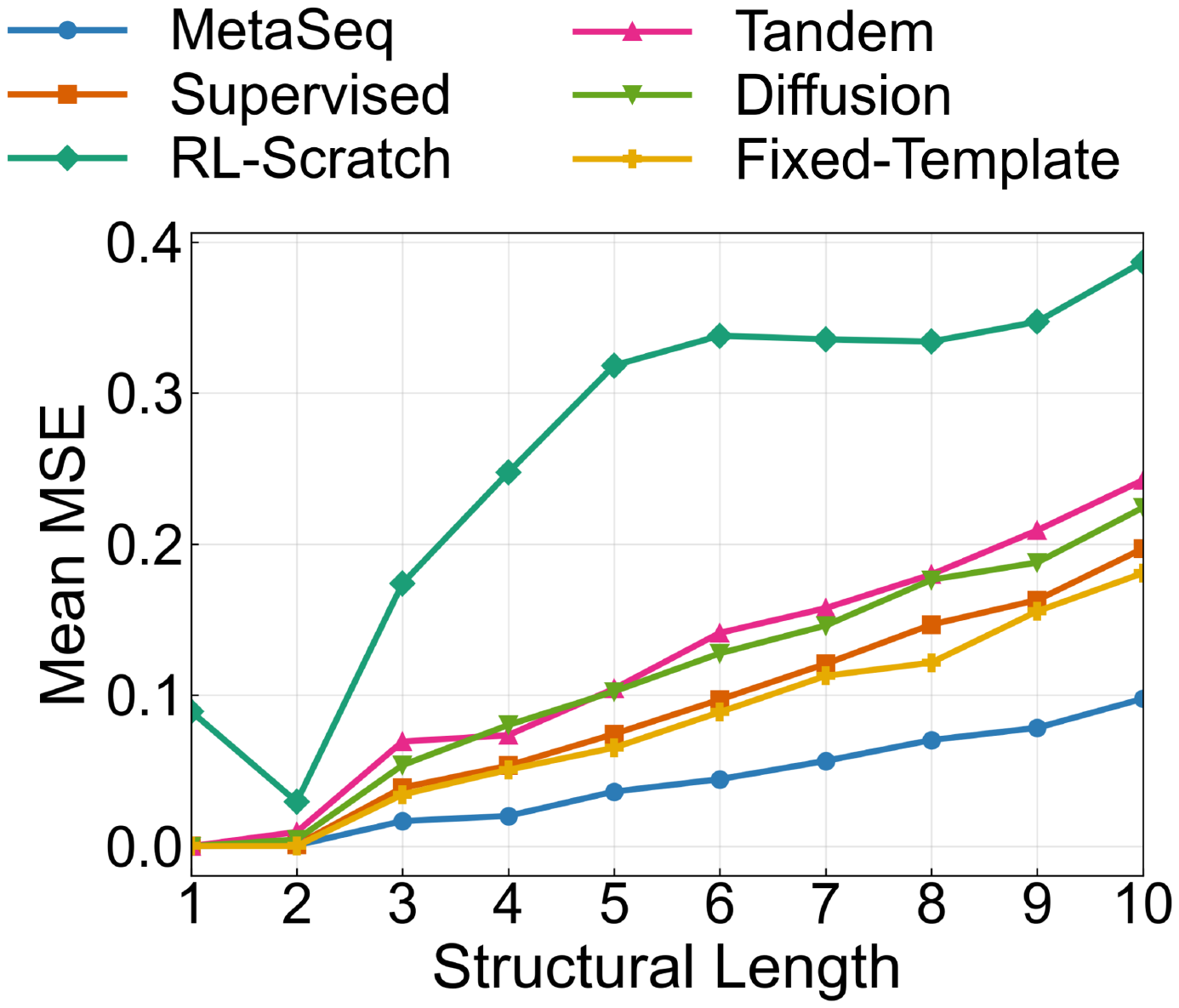}
		\label{fig:len_overall}
	}
	\vspace{-15pt}
	\caption{Overall Performance.} 
	\label{fig:overall}
	\vspace{-15pt}
\end{figure}


\subsection{Overall Performance}
\sssec{Method:} We evaluate the test set from low-frequency band, and compare  against five inverse-design baselines as described in Sec.~\ref{sec:eva_method}. All generated structures are then simulated in COMSOL, and the resulting responses are compared against the target responses. In addition to MSE, we also report Dynamic Time Warping (DTW), Pearson Correlation (PCC), and Cross-Correlation (XCorr) to assess response fidelity beyond point-wise error. These metrics shape similarity, linear correlation, and pattern alignment, respectively.

\sssec{Results:} Fig.~\ref{fig:CDF_overall} shows that  \sysname achieves the best overall performance among all baselines. Its gain over supervised-only   baseline highlights the value of response-level RL after supervised pretraining. \sysname's large margin over RL-Scratch further shows that stable initialization is critical for effective RL. The tandem baseline benefits from response-level supervision through a surrogate proxy, but cannot enforce structural validity because the validity checker is non-differentiable, which can favor invalid designs that better match the target. In contrast, \sysname incorporates validity-aware penalties during RL, enabling robust and physically-grounded optimization with better response matching. Fig.~\ref{fig:showcase} presents representative generated responses for diverse inverse-design objectives, including transmission responses with one, two, or three spectral peaks and wideband coverage, in both the high- and low-frequency bands. 

\begin{figure}[t]
	\centering
	\subfigure[CDF of MSE.]{
		\centering
		\includegraphics[width=0.465\linewidth]{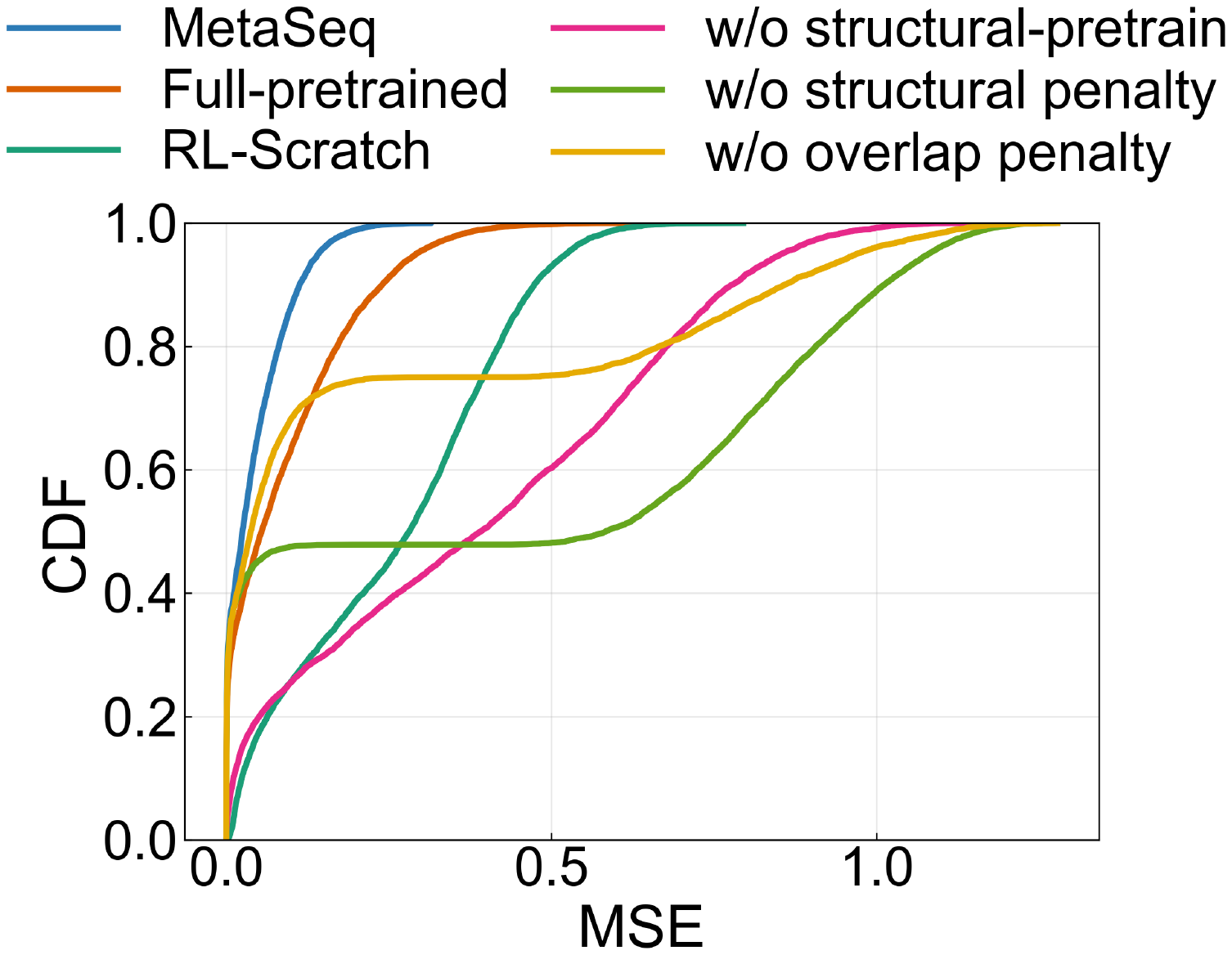}
		\label{fig:CDF_ablation}
	}\hspace{-10pt}
	\subfigure[MSE against structural length.]{
		\centering
		\includegraphics[width=0.465\linewidth]{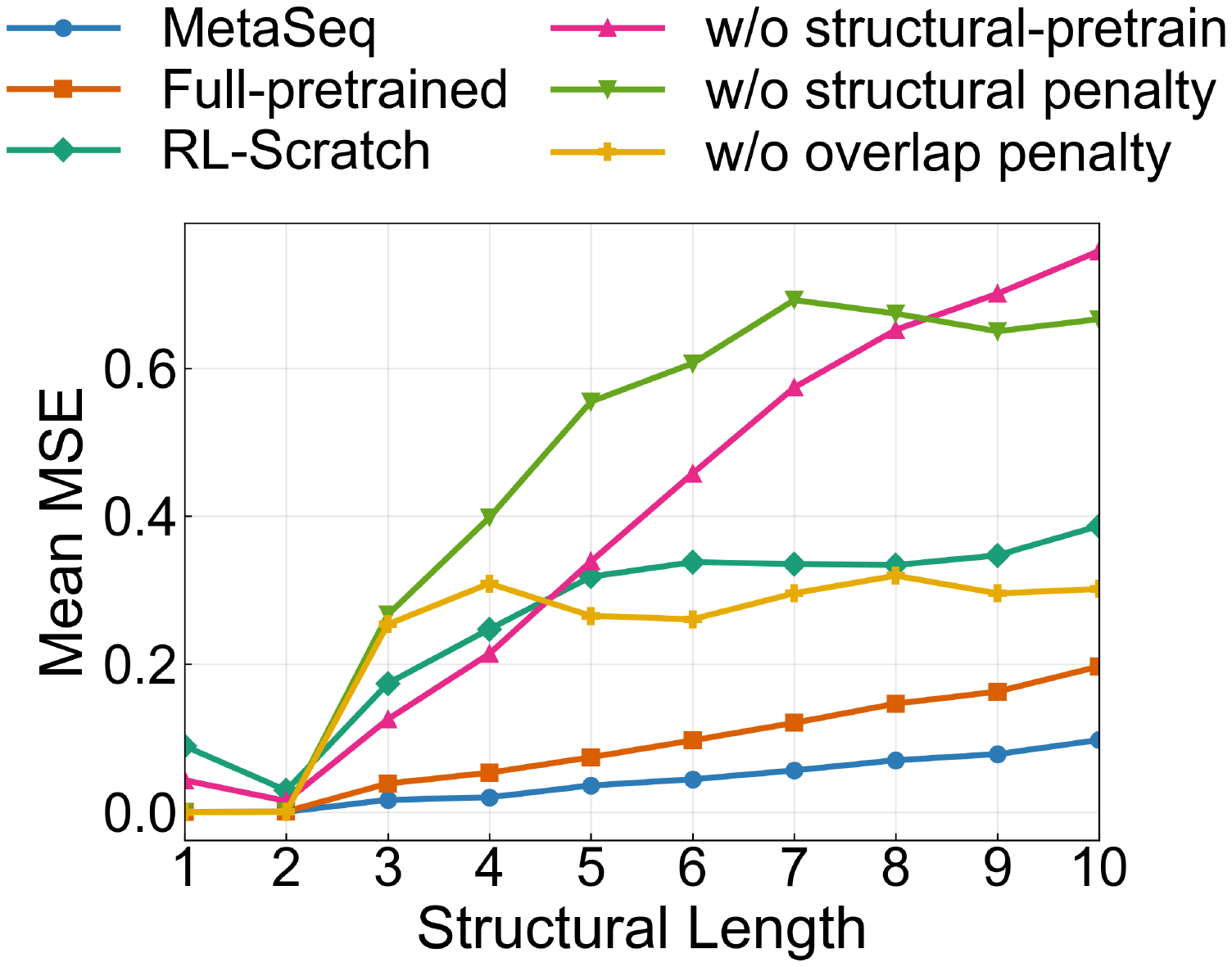}
		\label{fig:len_ablation}
	}
	\vspace{-15pt}
	\caption{Ablation Performance.} 
	\label{fig:ablation}
	\vspace{-16pt}
\end{figure}


Fig.~\ref{fig:len_overall} shows MSE vs. structural length (complexity). While errors rise with length across all methods, \sysname consistently achieves the lowest MSE. Notably, RL-Scratch struggles at short lengths due to initialization traps in constrained design spaces.


Table~\ref{tab:overall} summarizes the overall performance across multiple metrics, with the best and second-best results highlighted in green and red. \sysname consistently outperforms all baselines, achieving a 45\% reduction in MSE compared to the best baseline, Fixed-Template. The diffusion model yields an MSE of 0.1055 with a significant portion of this error ($0.0702$)  attributed to image compression. Although Fixed-Template achieves lower MSE than the supervised baseline, it yields worse DTW, indicating poorer peak alignment due to its topology restriction. In contrast, \sysname preserves better peak alignment while also offering fast inference.

\begin{table}[h]
	\definecolor{best}{HTML}{E2EFDA}    
	\definecolor{second}{HTML}{FCE4D6}  
	\renewcommand{\arraystretch}{1}
	\centering
	\caption{Overall Performance of \sysname.}
	\vspace{-10pt}
	\scalebox{0.65}{
		\begin{tabular}{lccccccc}
			\toprule[1.5pt]
			\textbf{Method} & MSE & DTW & PCC & XCorr & 
			\makecell[c]{Invalid\\Count} & 
			\makecell[c]{Training\\Time (min)} & 
			\makecell[c]{Infer\\ Time (sec)}\\
			\midrule
			
			Supervised & 0.0895 & \cellcolor{second}21.74 & 0.6644 & 150.87 & 8 & 119 & 0.028\\
			
			Diffusion & 0.1055 & 26.12 & 0.6348 & 149.31 & 11 & 14250 & 128\\
			
			Tandem & 0.1191 & 30.89 & 0.5888 & 148.29 & 11 & 173 & 0.028\\
			
			RL-Scratch & 0.2599 & 73.24 & 0.3589 & 118.70 & 0 & 78 & 0.039\\
			
			Fixed-Template & \cellcolor{second}0.0771 & 25.66 & \cellcolor{second}0.7412 & \cellcolor{second}154.25 & --- & --- & 18.55\\
			
			\textbf{\sysname} & \cellcolor{best}\textbf{0.0423} & \cellcolor{best}\textbf{16.26} & \cellcolor{best}\textbf{0.7737} & \cellcolor{best}\textbf{156.57} &\textbf{ 0} &\textbf{ 512} & \textbf{0.028}\\
			
			\bottomrule[1.5pt]
	\end{tabular}}
	\label{tab:overall}
	\vspace{-15pt}
\end{table}



\sssec{Invalid Count.} \sysname is designed to generate physically valid sequences that satisfy both structural validity and non-overlap constraints. Table.~\ref{tab:overall} shows that \sysname produces no invalid outputs, whereas several baselines still generate invalid designs. This shows that \sysname improves response matching without sacrificing structural correctness. Fixed-Template does not participate in this comparison because it selects candidates from the valid dataset.

\sssec{Training efficiency and inference time:} Training \sysname takes 512 minutes on 35k training samples with 7.5k validation samples,  including three phases: forward-model training for physical-solver proxy (47 min), supervised pretraining to converge (118 min), and  RL fine-tuning (347 min). Its inference time is only 0.028 seconds,  significantly faster than  diffusion and fixed-template.  Other methods have similar inference speed, but with lower  accuracy  than \sysname.

\

\subsection{Ablation Study}\label{sec:ablation}

\sysname uses  supervised pretraining to learn the structured language, then RL fine-tuning to explore alternative designs under response-based feedback. We now quantify the contribution of these design choices from Sec.~\ref{sec:generative} through ablation. 

\sssec{Method:}We consider five ablated variants: \textit{1) Full-pretrained:} removing the RL stage; \textit{2) RL-Scratch:} removing the supervised pretraining; \textit{ 3) w/o structural-pretrain:} which keeps the "pretrain + RL", but removes the structural-aware module including scalar reweighting and the bounded Sigmoid mapping; \textit{4) w/o structural penalty:} removing the structural penalty term $\lambda_{\mathrm{g}} \mathbb{I}_{\mathrm{g}}(\mathbf{d})$ in Eqn.~\ref{eq:rl_reward}; \textit{5) w/o overlap penalty:} removing the hinge loss term $\lambda_{\mathrm{ov}} \mathcal{L}_{\mathrm{hinge}} (\mathbf{d})$ in Eqn.~\ref{eq:rl_reward}. We compare  all variants with \sysname  using MSE and invalid count.


\sssec{Results:} As shown in Fig.~\ref{fig:CDF_ablation} and Fig.~\ref{fig:len_ablation}, removing either the structural or overlap penalty causes significant  degradation. The two constraints play complementary roles: the structure penalty ensures sequence-level integrity, while the overlap penalty enforces geometric feasibility. Removing the structural aware module also degrades performance as  structural length increases. This result suggests that weaker initialization in supervised pretraining cannot be fully corrected by later RL; thus, structure-aware pretraining is indispensable for stable policy improvement. 
 Compared with full-pretrained or RL from scratch, the results demonstrate that \sysname's superior performance stems from the synergy between structure-aware pretraining and validity-aware RL, rather than any single stage in isolation.

\sssec{Invalid Count:} The RL stage is pivotal in reducing invalid designs by explicitly penalizing structural and geometric violations. All models with the  validity-aware RL  reward (i.e., \sysname, RL-Scratch, and w/o structural-pretrain) produce 0 invalid cases. By contrast,  the pretrained model w/o the structural-aware module yields 93 invalid samples at convergence; while  RL later   corrects these invalid outputs, the poor initialization still limits fidelity on complex responses. The full-pretrained model produces 8  invalid cases, indicating that  supervised learning alone cannot ensure validity. Furthermore, removing individual penalties causes catastrophic failures:  w/o structural penalty and w/o overlap penalty produce 3,912 and 1,874 invalid outputs, respectively, underscoring the necessity of explicit constraints in the RL reward.
\begin{figure}[t]
	\centering
	\includegraphics[width=0.95\linewidth]{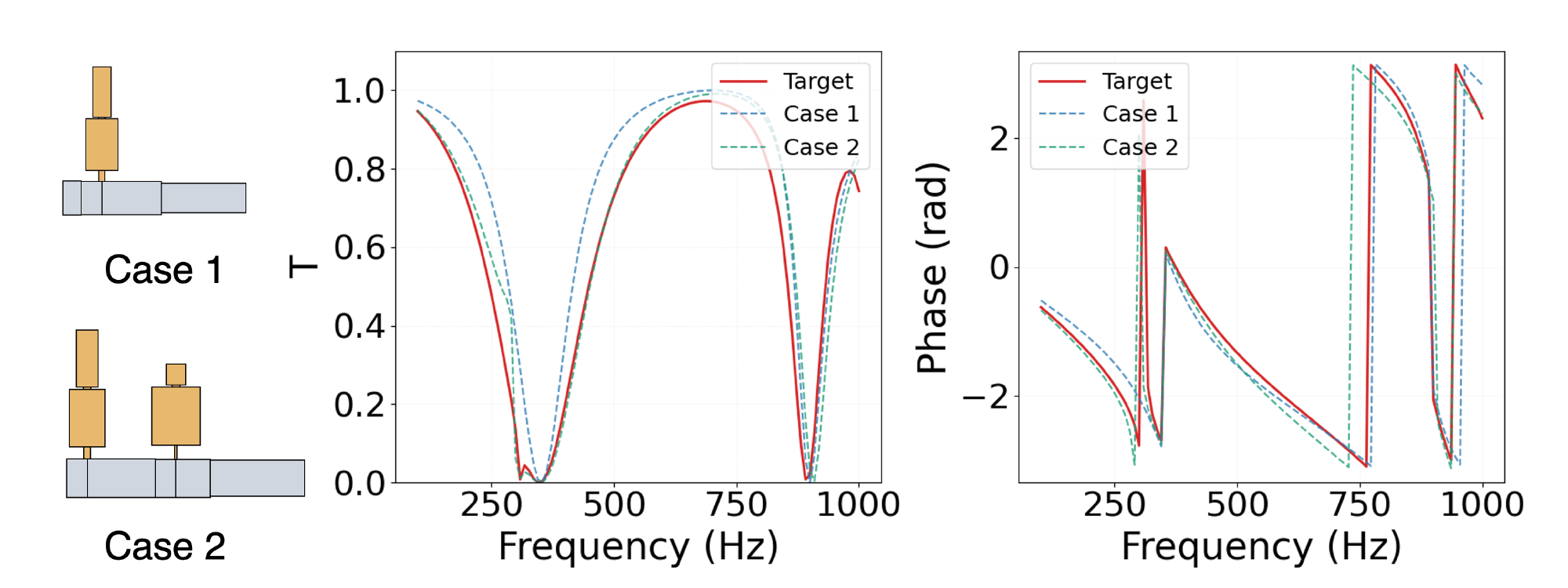}
	\vspace{-12pt}
	\caption{Structural Diversity.} 
	\label{fig:diverse}
	\vspace{-15pt}
\end{figure}

\subsection{Structural Diversity}
A key benefit of \sysname's RL-based response-matching objective is that it better accommodates the one-to-many nature of inverse design by encouraging multiple distinct structures for the same target response, rather than overfitting to a single reference design, thereby better supporting downstream needs such as fabrication robustness or material usage. As shown in Fig.~\ref{fig:diverse}, the two generated designs achieve similarly matched responses despite having markedly different layouts. Notably, Case 1 is shorter and smaller than Case 2, suggesting a more material-efficient alternative.

\begin{figure*}[t]
	\centering
	\addtocounter{subfigure}{-1}
	\subfigure{
		\includegraphics[width=0.9\linewidth]{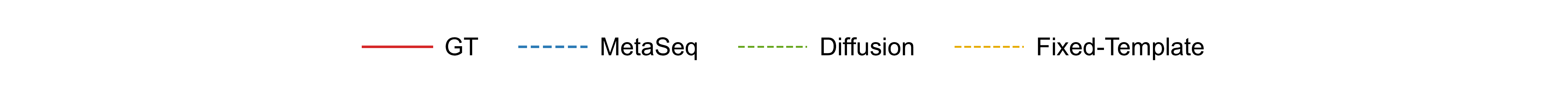}
	}\vspace{-20pt}
	\subfigure[One spectral peak in T (100-1kHz)]{
		\centering
		\includegraphics[width=0.155\linewidth]{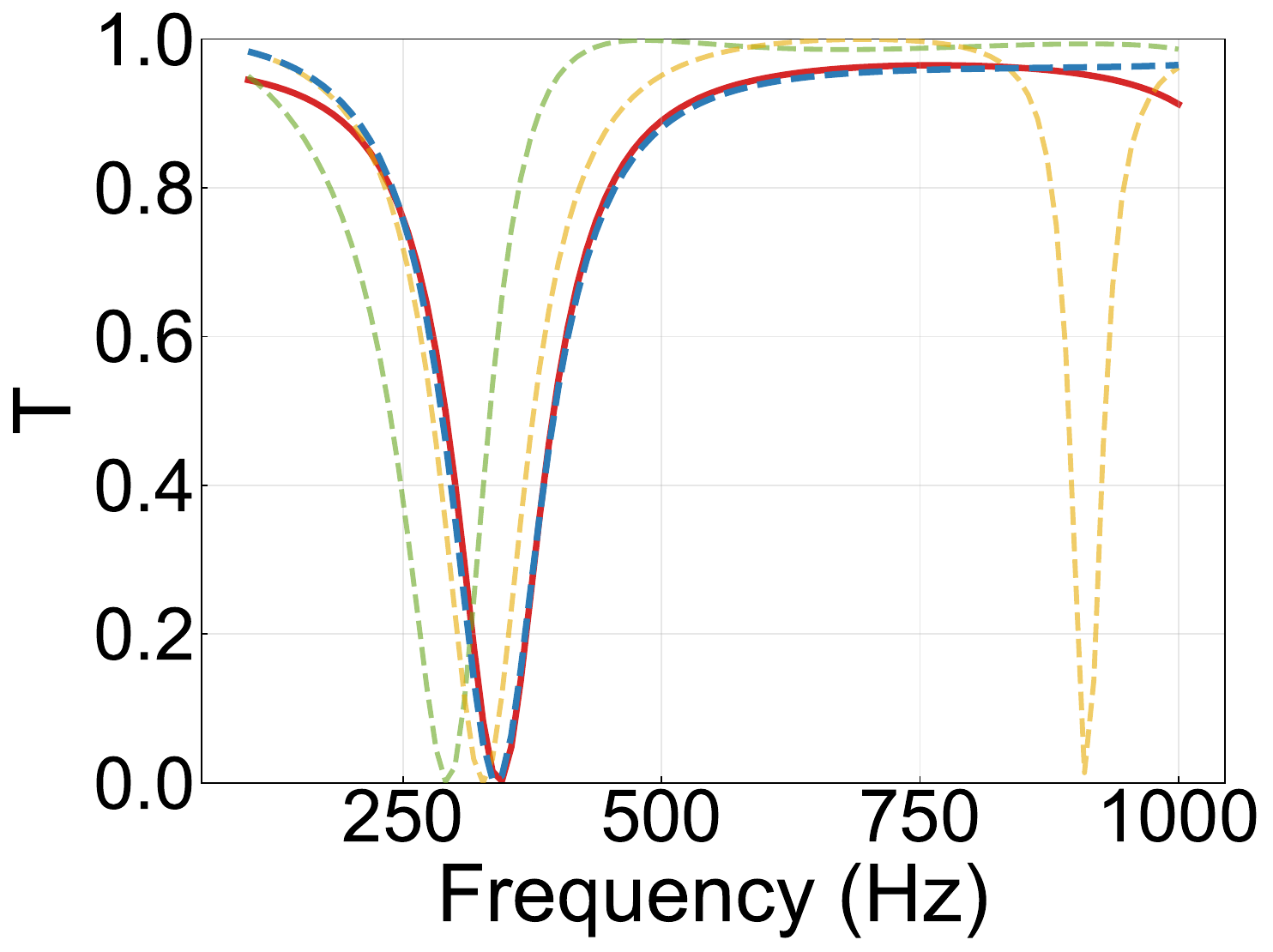}
		\includegraphics[width=0.155\linewidth]{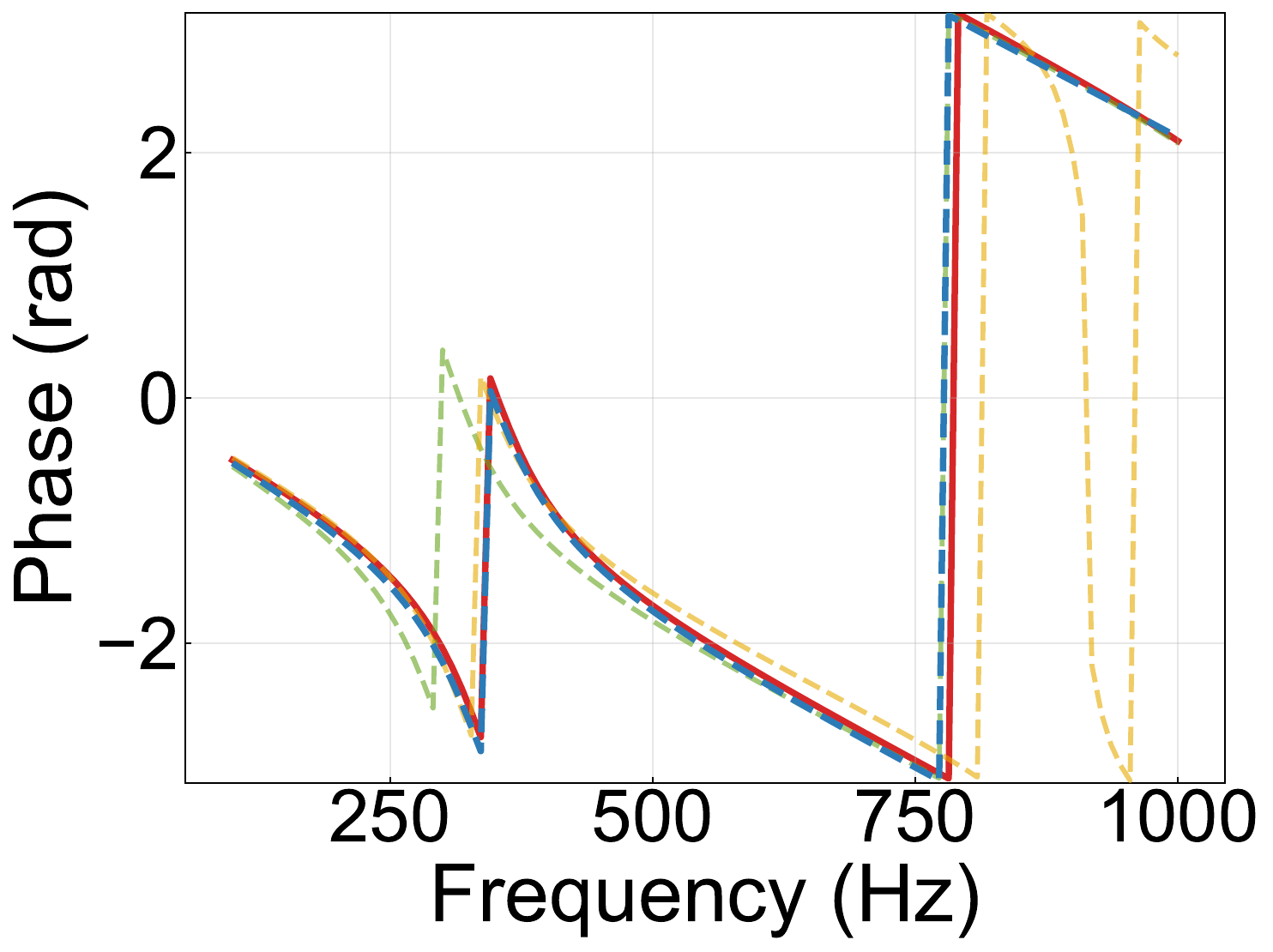}
		\label{fig:low_single}
	}\hfill
	\subfigure[Two spectral peaks in T (100-1kHz)]{
		\centering
		\includegraphics[width=0.155\linewidth]{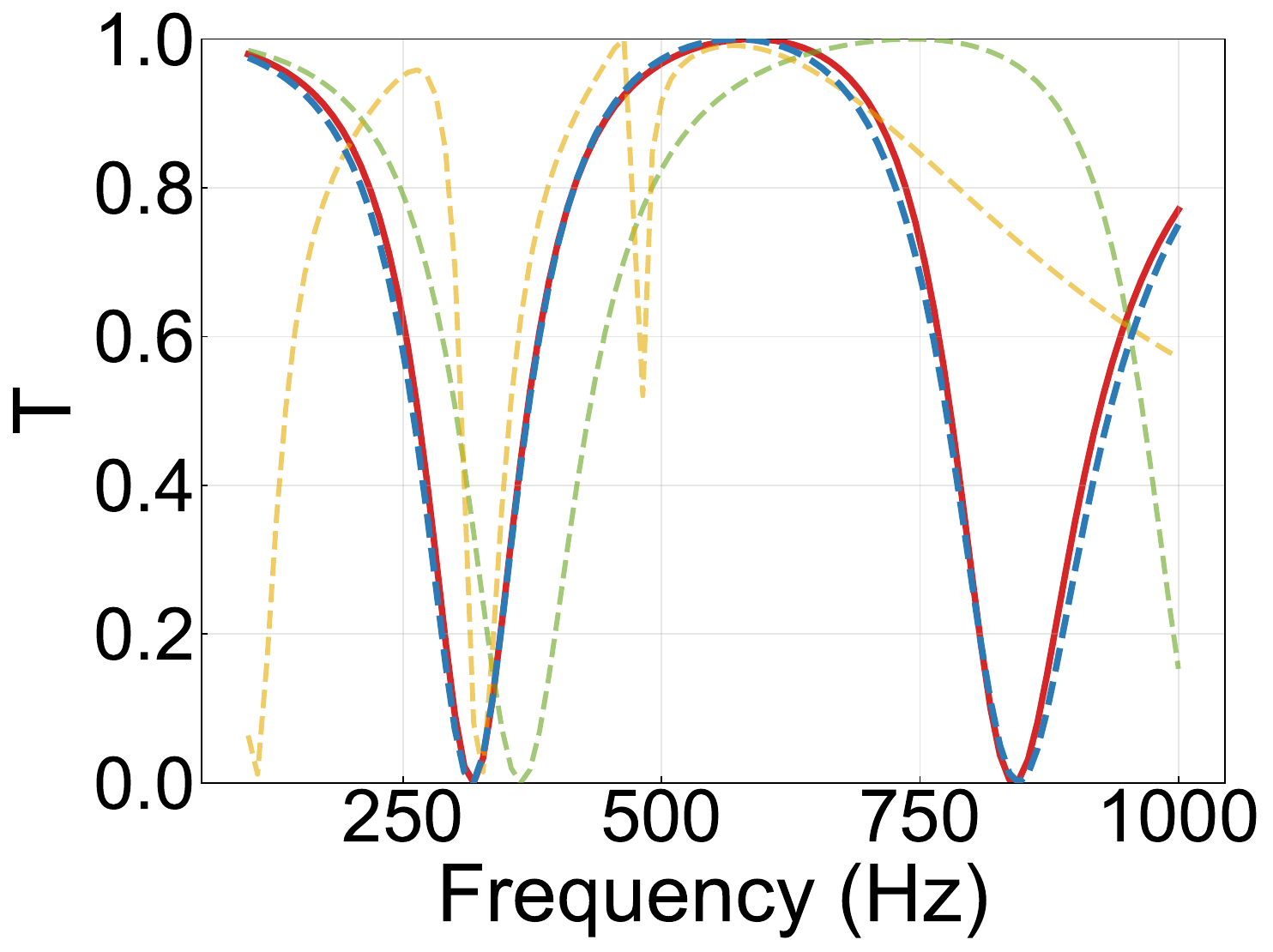}
		\includegraphics[width=0.155\linewidth]{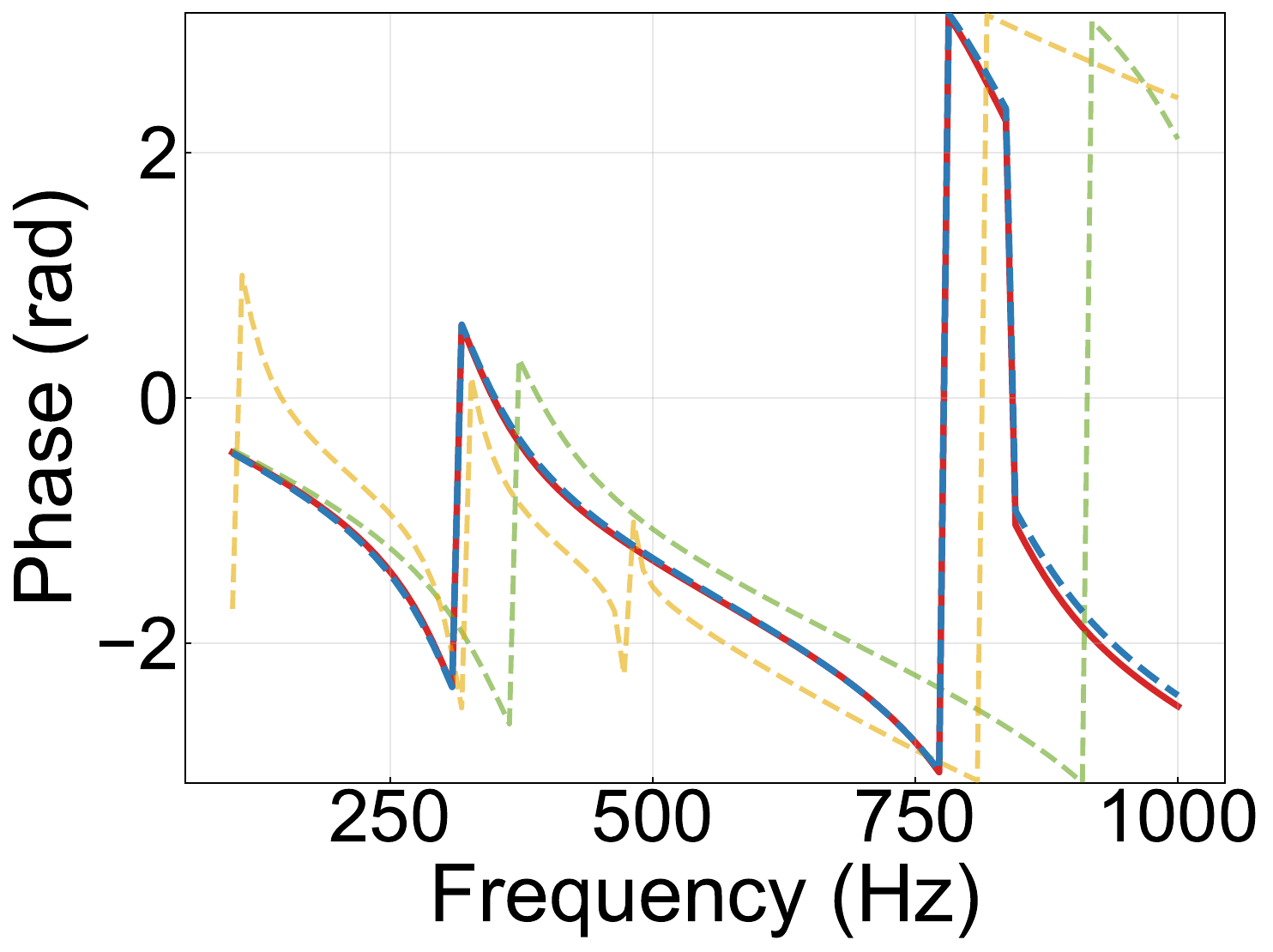}
		\label{fig:low_dual}
	}\hfill
	\subfigure[Three spectral peaks in T (100-1kHz)]{
		\centering
		\includegraphics[width=0.155\linewidth]{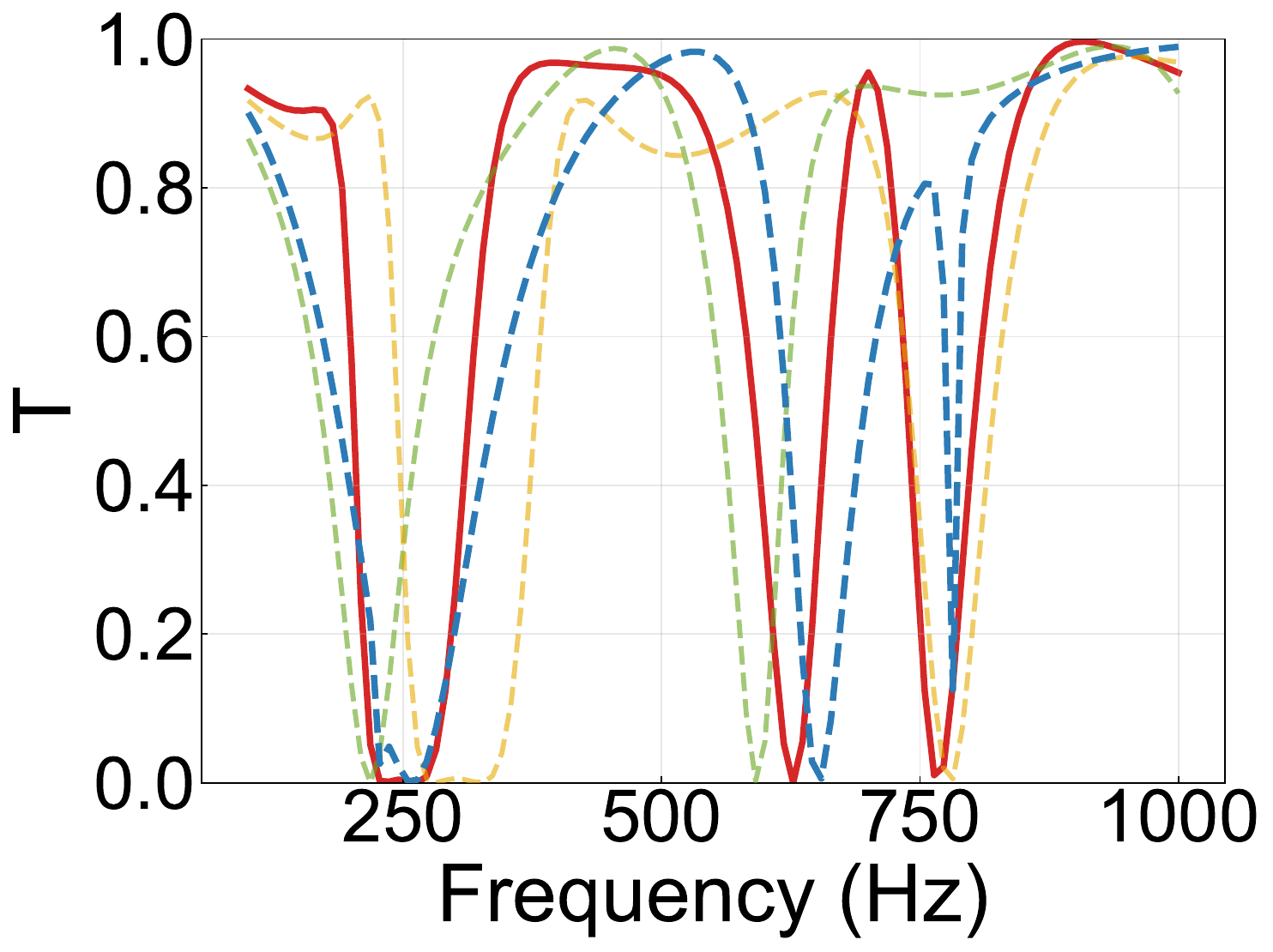}
		\includegraphics[width=0.155\linewidth]{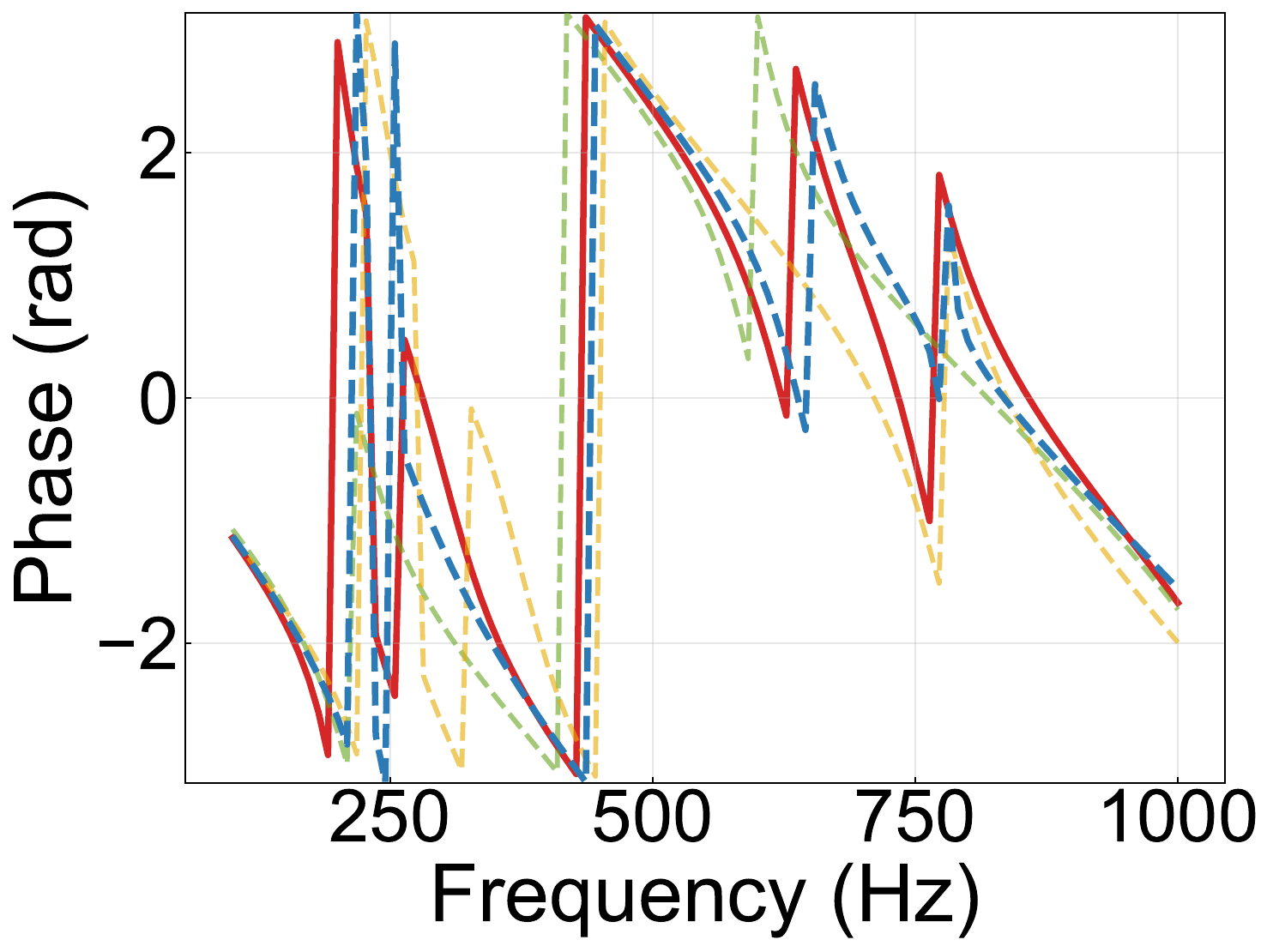}
		\label{fig:low_triple}
	}\vspace{-10pt}
	
	\subfigure[Example 1 (16-24kHz)]{
		\centering
		\includegraphics[width=0.155\linewidth]{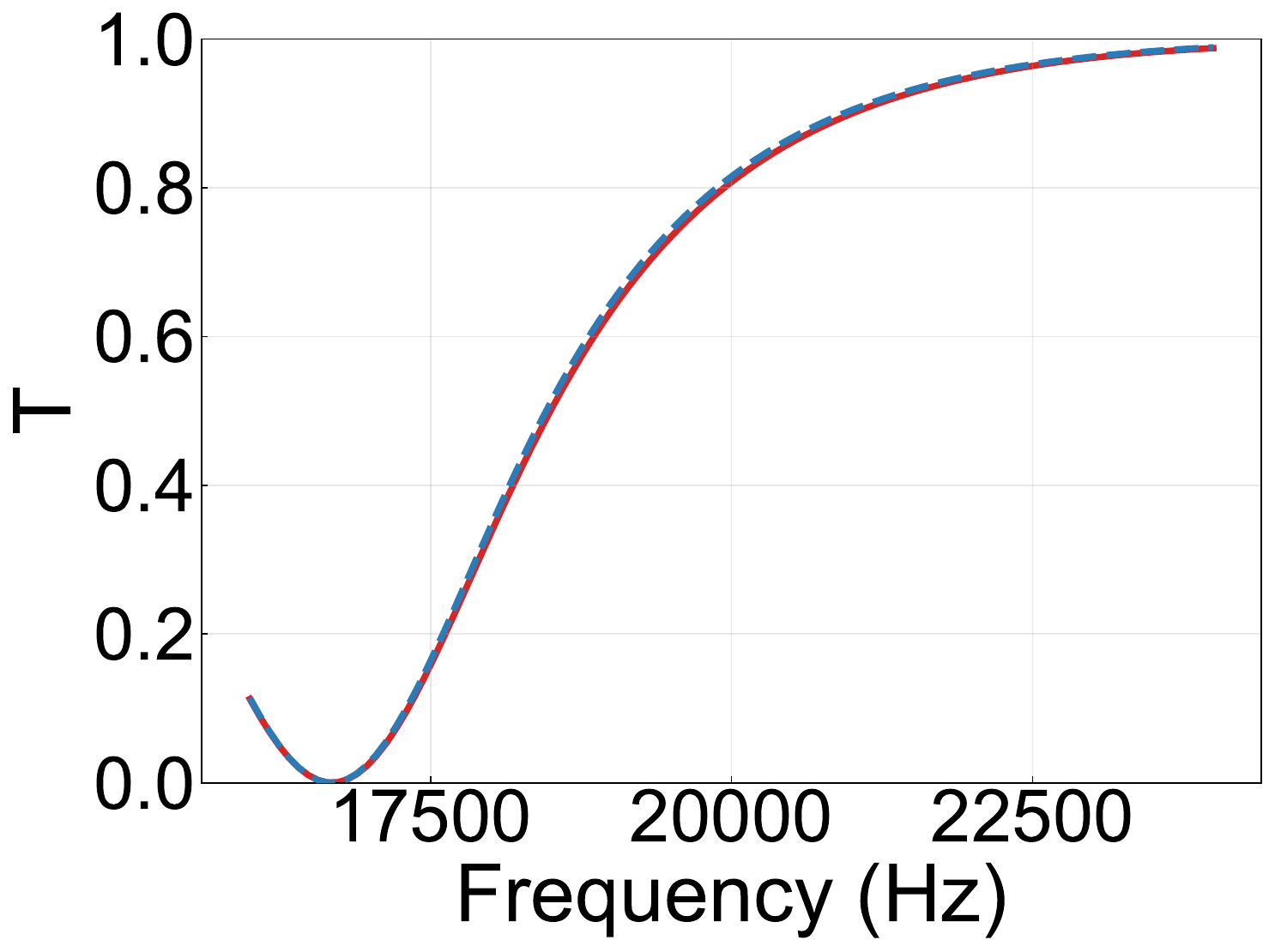}
		\includegraphics[width=0.155\linewidth]{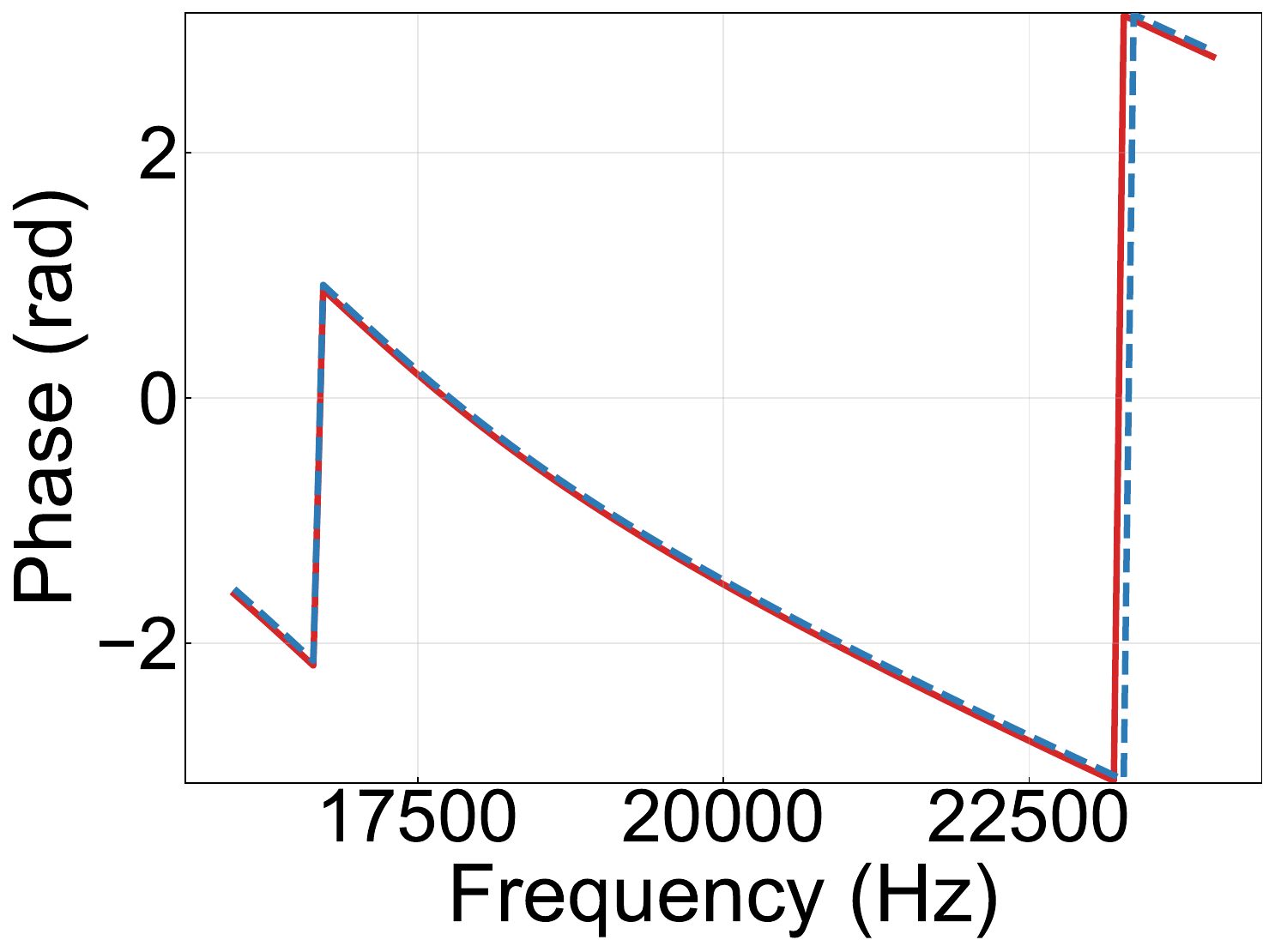}
		\label{fig:high_single}
	}\hfill
	\subfigure[Example 2 (16-24kHz)]{
		\centering
		\includegraphics[width=0.155\linewidth]{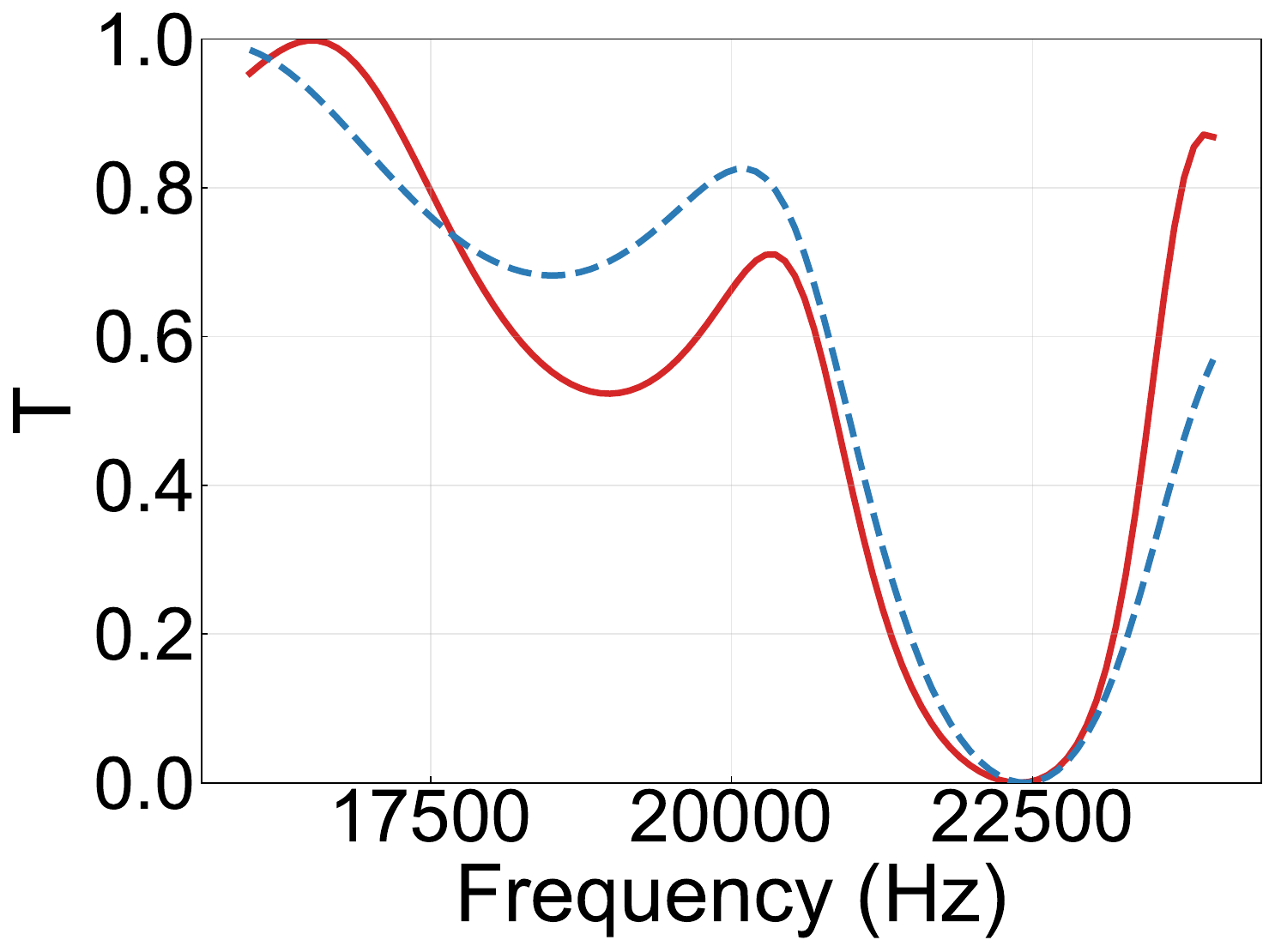}
		\includegraphics[width=0.155\linewidth]{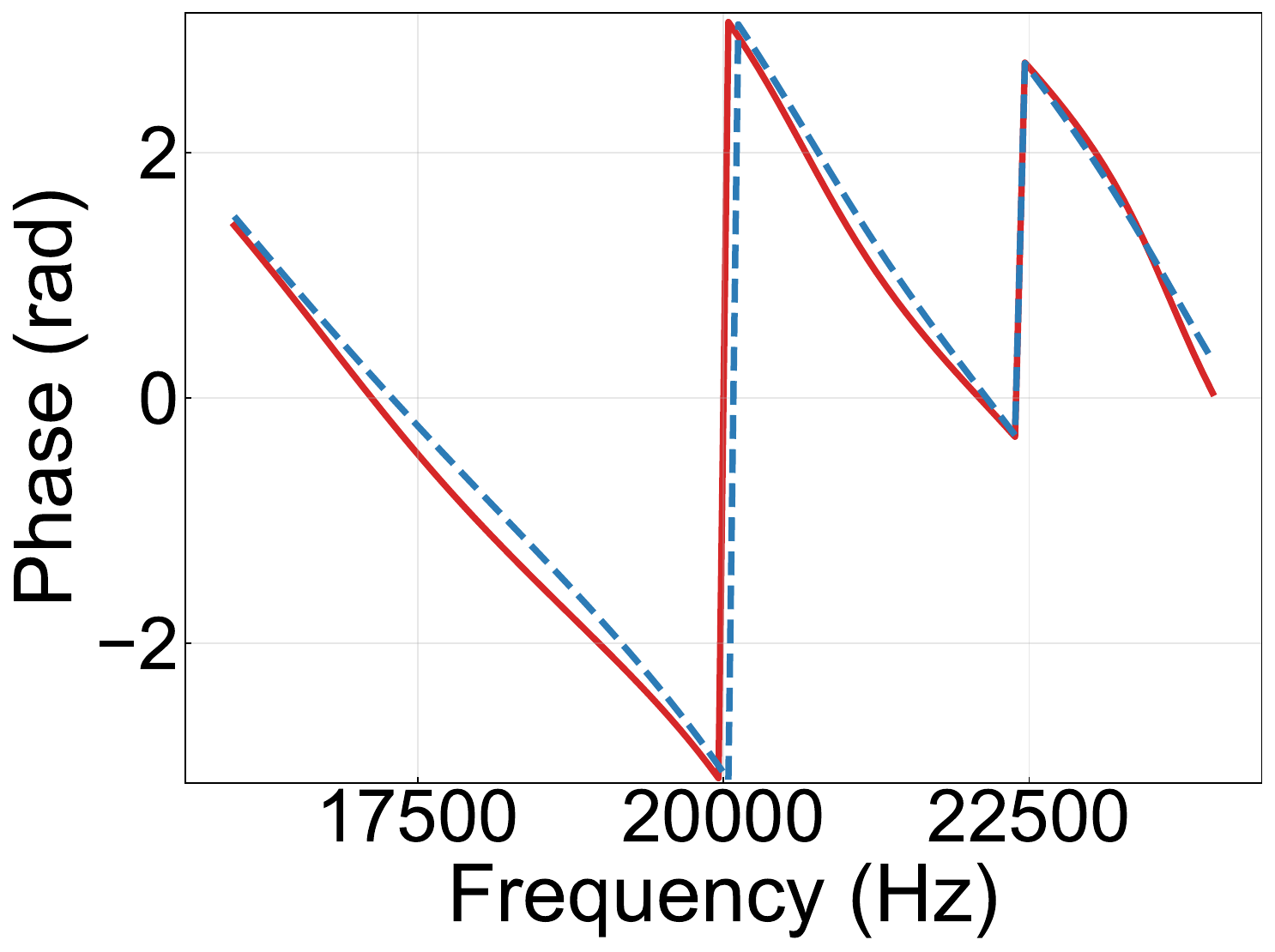}
		\label{fig:high_dual}
	}\hfill
	\subfigure[Example 3 (16-24kHz)]{
		\centering
		\includegraphics[width=0.155\linewidth]{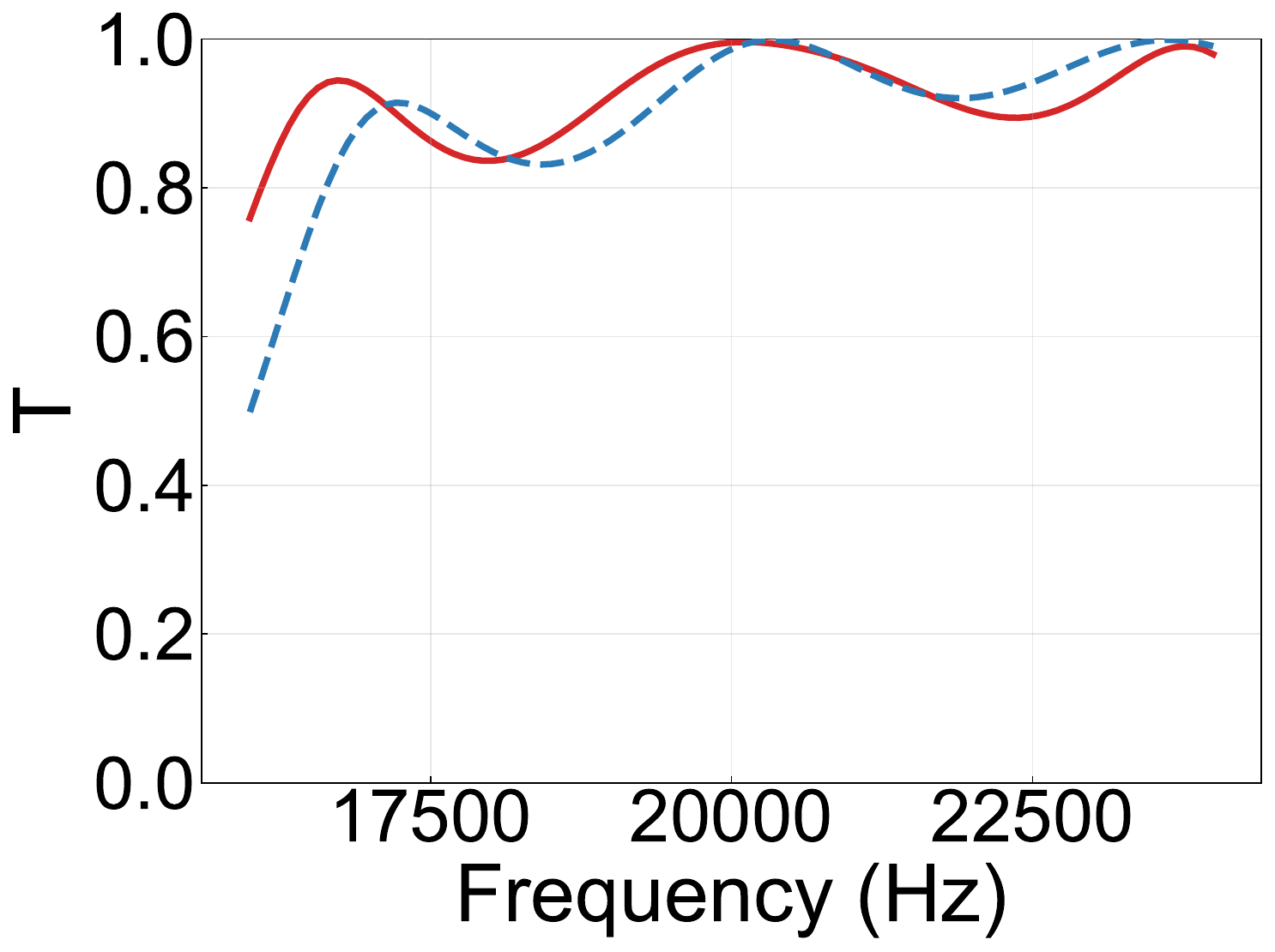}
		\includegraphics[width=0.155\linewidth]{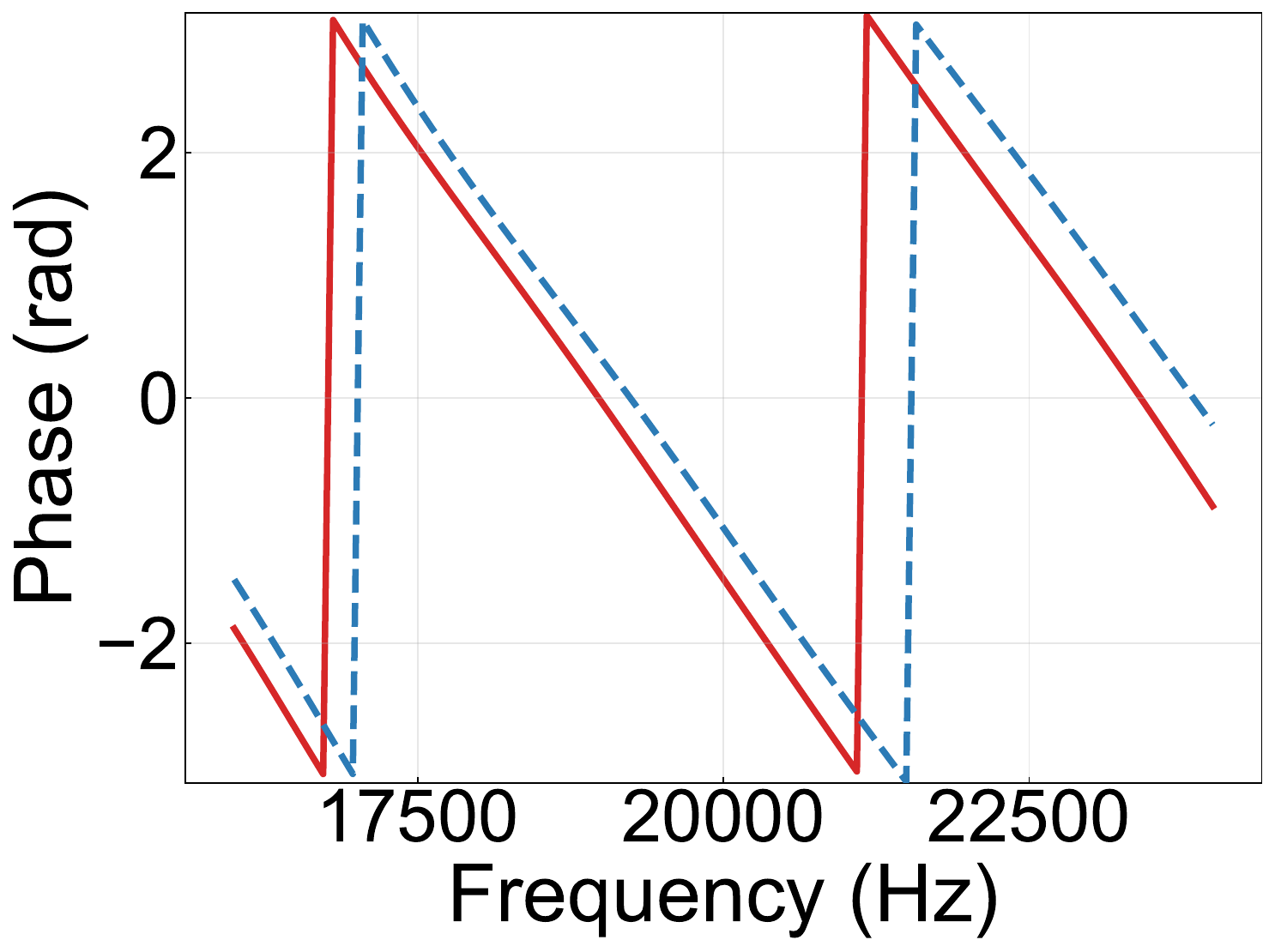}
		\label{fig:high_triple}
	}
	\vspace{-15pt}
	\caption{Representative target and generated responses of \sysname in low- and high-frequency bands.} 
	\label{fig:showcase}
	\vspace{-10pt}
\end{figure*}

%
%
\section{Related Work}
\sssec{Acoustic Metamaterials.} AMMs~\cite{fok2008acoustic} are sub-wavelength units composed of cavities and ducts that manipulate acoustic amplitude and phase. They serve as standalone devices for wave control tasks such as sound absorption, wavefront engineering~\cite{Ma2014NatMater, LiAssouar2016APL, Assouar2018NRM, Bok2018PRL}. AMMs have also been integrated into larger acoustic metasurfaces~\cite{Assouar2018NRM} for applications including sensing~\cite{wang2025metasonic,fu2024pushing,fu2024adaptive} and beam steering~\cite{xie2014wavefront, li2024mudis, zhou2025m2silent}.
Realizing such metasurfaces requires customizing individual AMM units with precise responses, including broadband  behaviors~\cite{tang2021broadband, chen2018broadband, ning2025metaguardian,xie2014wavefront}.

\sssec{Acoustic Metamaterial Inverse Design.}  Conventional inverse design is often formulated as  iterative optimization. Automated methods like Genetic Algorithms~\cite{jafar2018adaptive,johnson1997genetic,Li2012JASAGA} and gradient-based methods~\cite{jensen2011topology,AppliedAcoustics2023TOAMS,noguchi2021topology} improve over manual heuristics~\cite{johnson1997genetic, jensen2011topology}, but remain slow, prone to poor local optima, and dependent on  predefined parameterization or structural templates~\cite{Molesky2018NatPhotonInverseDesign}. Deep learning has also been used to regress geometric parameters based on target responses~\cite{Zhang2023ActaMechSinInverseAcoustic,Lv2024MaterialsAcousticMetasurfaceDL,Zhou2025SciRepBroadbandAbsDL,Zhu2025ActaAcusticaANNAbsorber,AppliedAcoustics2025CylPlateDL,AppliedAcoustics2026NewInvDL,JSV2025InvDL118789}. While more efficient, these methods likewise assume a fixed template, which limits their ability to realize broadband responses.
To move beyond predefined templates, generative models such as GANs and CVAEs have been explored~\cite{an2021multifunctional, Yan2025YMSSP}. While diffusion models show promise in electromagnetic domains~\cite{zhao2025metagen}, their use in acoustics remains limited. First, acoustic AMMs require  continuous fluid paths, while  pixel-based generation
can  produce disconnected voids. Second, they suffer from a tradeoff between geometric precision and computational cost:  high-resolution image-based training is  expensive, compressed images sacrifice the fine structural details required for accurate response matching.

\sssec{$\blacksquare$ Summary of Differences.} In contrast, \sysname (1) represents AMM units as structured sequences  to preserve geometric fidelity; (2) enforces an {explicit topological grammar} to guarantee continuous fluid paths; and (3) combines {reinforcement learning with a physics-based solver} to generate multiple valid structures beyond fixed templates.



\section{Discussion}

\sssec{Obtaining the Target Response:} This paper focuses on inverse design for a given target response of an AMM structure. That  target can be specified differently across applications. For sound absorption, it can be specified directly as a desired amplitude-phase response over frequency. For beam focusing, where an AMM unit serves as a building block of a larger metasurface, each unit's target can instead be derived from system-level design using traditional analysis~\cite{yu2014flat} or LLM-assisted pipelines such as~\cite{zhao2025metagen}, where the LLM generates code for the underlying analysis.

\sssec{Extensibility of Structural Vocabulary:}
The current vocabulary covers common acoustic primitives such as ducts, necks, and cavities. However, the framework can be extended to richer components, including micro-perforated plates, elastic membranes, and poro-elastic layers~\cite{fok2008acoustic}. This extension only requires adding new tokens and their  analytical models (e.g., transfer matrices) to the physical  solver; the  neural architecture remains unchanged. 



\sssec{Extending to Environment Coupling:}
In the current implementation, we do not explicitly model coupling with the surrounding environment. 
However, when environmental effects are known,  they can be incorporated into the physical solver in Appendix~\ref{app:solver_details}, e.g., as an additional transfer block in the cascaded matrix $\mathbf{M}$ to account for coupling between the structure and the external load.

\sssec{Re-Training Conditions:}  The pretrained model requires retraining when the underlying design space changes, such as with new structural primitives, different geometric parameter ranges, or target frequencies far outside the original training distribution. RL fine-tuning can instead be rerun separately when the optimization objective or reward changes, for example to support coupling with a specific environment.

\sssec{Physics-Solver Calibration:} The correction mechanism for the physical solver will be revisited  when new primitives are introduced. For instance, the current implementation assumes 2D primitives, and extending the framework to 3D structures would require a new calibration.

\section{Conclusion}
This paper presents \sysname, a physics-guided generative framework for acoustic metamaterial inverse design that targets broadband acoustic responses. \sysname first introduces a novel structured language for AMM structures and constructs a balanced, high-fidelity dataset.  Built on this representation, \sysname casts inverse design as a sequence-to-sequence generation task and learns to generate multiple valid structures with low response error. Detailed evaluation  shows that \sysname improves accuracy and efficiency over existing template-based, image-based methods.


\bibliographystyle{plain}
\bibliography{reference}


\appendix
\section{Overlap Checker}\label{app:checker}
The overlap checker is a geometric validation algorithm (Algo.~\ref{alg:geom_validity_core}) designed to verify the spatial feasibility of a structured sequence. It ensures that a given topology can be physically manufactured without spatial conflicts. The algorithm first reconstructs the geometry, which iterates through tokens such as duct, shunt, and parallel blocks. It dynamically manages a local coordinate system, which tracks the anchors and heights to map the sequence to a 2D bounding box layout. Following reconstruction, the algorithm performs a Collision Detection pass across each box pairs to identify any spatial conflicts and calculate the total overlap area. The output includes a binary isValid flag and the quantitative overlap metric, which is not only used for data construction, but also for serving as a penalty term for reward model in RL workflows.

\begin{algorithm}[t]
	\setstretch{0.5}
	\caption{\textbf{Overlap Checker}}
	\label{alg:geom_validity_core}
	\KwIn{\texttt{token\_ids}, \texttt{params} in \texttt{sequence}}
	\KwOut{\texttt{isValid}, \texttt{overlapArea}}

	\textbf{\# Initialization} \\
	\texttt{boxes} $\leftarrow [\,]$; \texttt{totalOverlap} $\leftarrow 0$ \;
	\texttt{cur\_x, cur\_y} $\leftarrow 0, 0$; \texttt{para\_anchor\_y} $\leftarrow 0$; \texttt{max\_para\_h} $\leftarrow 0$ \;
	\texttt{isValid} $\leftarrow \texttt{True}$ \;
	
	\textbf{\# 1. Geometry Reconstruction} \\
	\For{\texttt{token, p} in \texttt{sequence}}{
		\Case{\texttt{Duct}}{
			Place duct at \texttt{cur\_x}; \texttt{cur\_x} $\leftarrow$ \texttt{cur\_x} + \texttt{p.length} \;
		}
		\Case{\texttt{Shunt\_Start}}{
			\texttt{cur\_y} $\leftarrow$ \texttt{duct\_wall\_height} \tcp{Move to side branch}
		}
		\Case{\texttt{Para\_Start}}{
			\texttt{para\_anchor\_y} $\leftarrow \texttt{cur\_y}$; \texttt{max\_para\_h} $\leftarrow 0$ \tcp{Set anchor}
		}
		\Case{\texttt{primitive} (Inside Para)}{
			Place box at (\texttt{cur\_x}, \texttt{para\_anchor\_y}) \;
			\texttt{max\_para\_h} $\leftarrow \max(\texttt{max\_para\_h}, \texttt{p.height}) $ \;
		}
		\Case{\texttt{Para\_End}}{
			\texttt{cur\_y} $\leftarrow \texttt{para\_anchor\_y} + \texttt{max\_para\_h}$ \tcp{Jump to end of block}
		}
		\Case{\texttt{primitive} (Inside Shunt)}{
			Place box at (\texttt{cur\_x}, \texttt{cur\_y}); \texttt{cur\_y} $\leftarrow \texttt{cur\_y} + \texttt{p.height}$ \;
		}
		\Case{\texttt{Shunt\_End}}{
			\texttt{cur\_y} $\leftarrow 0$ \tcp{Return to main path}
		}
	}
	
	\textbf{\# 2. Collision Detection} \\
	\ForEach{pair $(B_i, B_j)$ in \texttt{boxes}}{
		\If{$B_i \cap B_j \neq \emptyset$}{
			\texttt{totalOverlap} $\leftarrow \texttt{totalOverlap} + \mathrm{Area}(B_i \cap B_j)$ \;
			\texttt{isValid} $\leftarrow \texttt{False}$ \;
		}
	}
	\Return{\texttt{isValid}, \texttt{totalOverlap}} \;
\end{algorithm}

\section{Physical Solver Details}\label{app:solver_details}
\begin{figure}[t]
	\centering
	\includegraphics[width=0.68\linewidth]{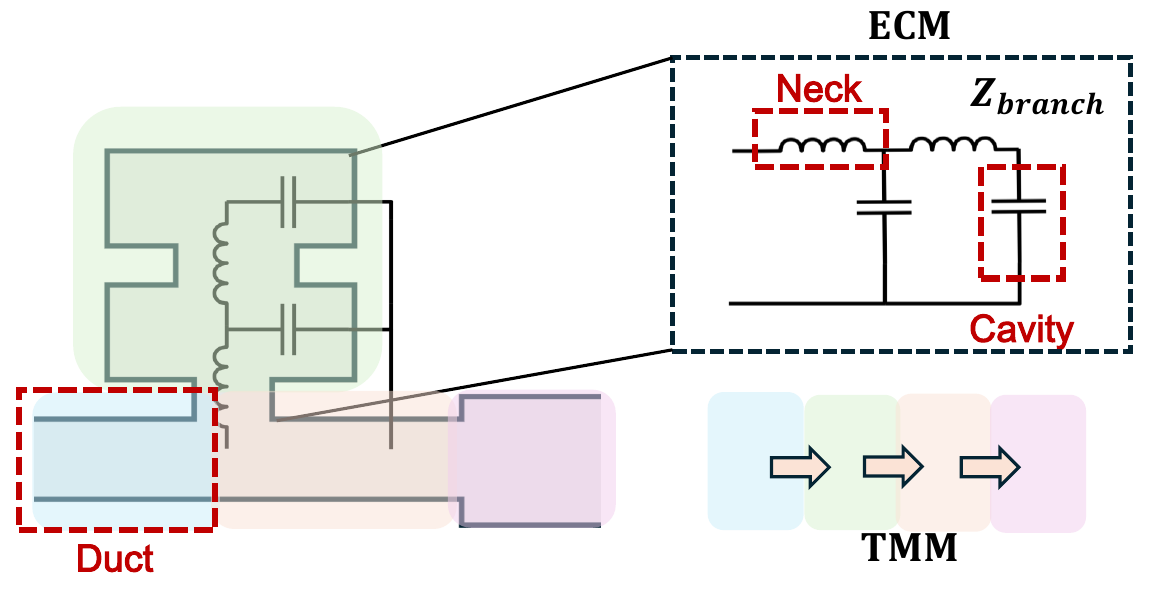}
	\vspace{-10pt}
	\caption{Illustration of physical solver for acoustics.} 
	\label{fig:TMM}
	\vspace{-10pt}
\end{figure}

This section provides the detailed formulation omitted from Sec.~\ref{sec:solver}. We divide the acoustic primitives into resonant units located in the branches and propagation units located in the main path. As shown in Fig.~\ref{fig:TMM}, for the resonant unit, we can use equivalent circuit calculations; while for the main circuit unit, we establish a transfer matrix analysis as following.


\sssec{Primitive-Level Modeling.}
We model each decomposed primitive using either a lumped-element or distributed-wave approximation.
\begin{itemize}[leftmargin=10pt,topsep=0pt]
	\item \textit{Lumped elements:} for deep sub-wavelength resonant parts,
	\[
	Z_{neck} \approx j\omega\,\rho_0\,\frac{l_{neck}}{S_{neck}},
	\qquad
	Z_{cav} \approx \frac{1}{j\omega C_a},\quad C_a=\frac{V_{cav}}{\rho_0 c_0^2}.
	\]
	\item \textit{Distributed elements:} for propagation-dominant ducts. 
	\[
	k=\frac{\omega}{c_0},\qquad Z_c=\frac{\rho_0 c_0}{S},
	\qquad Z_{duct}=-jZ_c\cot(kl_{eff}).
	\]
\end{itemize}
Specifically, coiled structures (denoted by \texttt{[Coil]}) can be conceptually unrolled into equivalent straight pipes~\cite{liang2012extreme}, exhibiting the same physical properties as a duct. Notably, the effective length of a coiling unit is usually much longer than its physical thickness, which enables wide-range phase modulation.

\sssec{Topology Composition Rules.}
For parsed branch topologies, the equivalent branch input impedance $Z_b$ is computed recursively:
\[
Z_b^{(series)}=\sum_i Z_i,
\qquad
Z_b^{(parallel)}=\left(\sum_i\frac{1}{Z_i}\right)^{-1}.
\]
On the main path, we use transfer matrices for two connection types:
\[
\mathbf{M}_{shunt}=\begin{bmatrix}1 & 0\\ 1/Z_b & 1\end{bmatrix},
\qquad
\mathbf{M}_{series}=\begin{bmatrix}
	\cos(kl) & jZ_c\sin(kl)\\
	\frac{j}{Z_c}\sin(kl) & \cos(kl)
\end{bmatrix}.
\]

\sssec{Global Response Extraction.}
By cascading all parsed elements along the main path,
\[
\mathbf{M}=\prod_i \mathbf{M}_i
=\begin{bmatrix}M_{11} & M_{12}\\ M_{21} & M_{22}\end{bmatrix}.
\]
With port impedance $Z_0=\rho_0c_0/S_0$, the complex transmission and reflection coefficients are:
\[
t=\frac{2Z_0}{M_{11}Z_0+M_{12}+M_{21}Z_0^2+M_{22}Z_0},
\]
\[
r=\frac{M_{11}Z_0+M_{12}-M_{21}Z_0^2-M_{22}Z_0}{M_{11}Z_0+M_{12}+M_{21}Z_0^2+M_{22}Z_0}.
\]
Finally, we compute the response profiles as
\[
T=|t|^2, R=|r|^2, \\
\phi=\arg(t)\;\text{(or }\arg(r)\text{ for reflection tasks)}.
\]

\sssec{Adaptation for Reflection Tasks.}
To adapt the solver for reflection-based meta-atoms (e.g., with a rigid backing), the primitives are parsed in sequence from the incident port to the termination boundary. In this configuration, the exit port is replaced by a rigid wall condition ($Z_L \to \infty$), and the global input impedance is simplified to $Z_{in} = M_{11}/M_{21}$. The complex reflection coefficient is then determined as $r = (Z_{in} - Z_0)/(Z_{in} + Z_0)$. This approach ensures that the primitive ordering remains physically consistent with the inward propagation path while the boundary condition implicitly accounts for the phase accumulation of the reflected wave. Notably, this will not affect our language; we only need to modify the physical solver accordingly.

\section{ Fidelity Analysis for the Calibrated Solver}\label{app:fidelity}

\begin{figure}[t]
	\centering
	\subfigure[MSE and optimization time cost across the hold-out dataset under impact of the number of  FEM samples.]{
		\centering
		\includegraphics[width=0.46\linewidth]{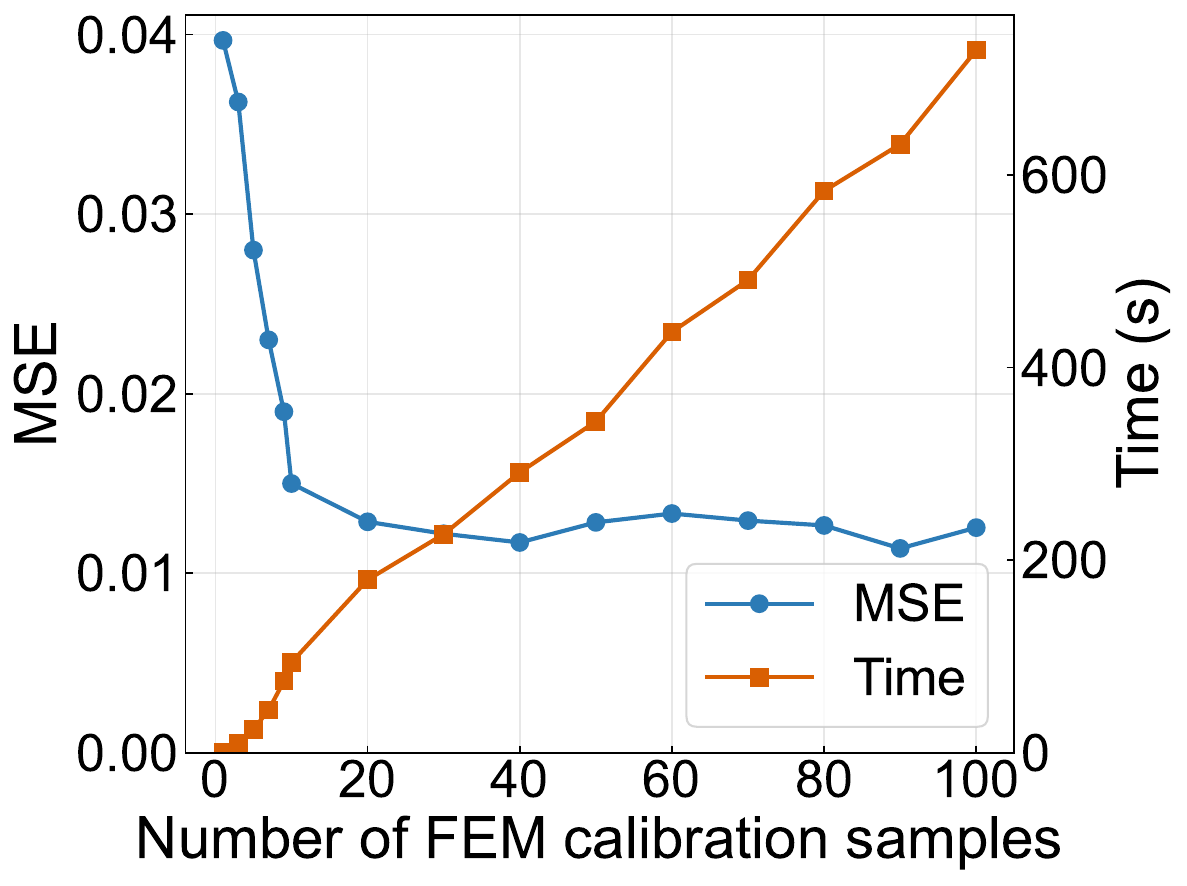}
		\label{fig:MSE_time_FEM}
	}\hspace{5pt}
	\subfigure[CDF comparison of MSE. (\sysname, empirical, no correction, and random sampling)]{
		\centering
		\includegraphics[width=0.46\linewidth]{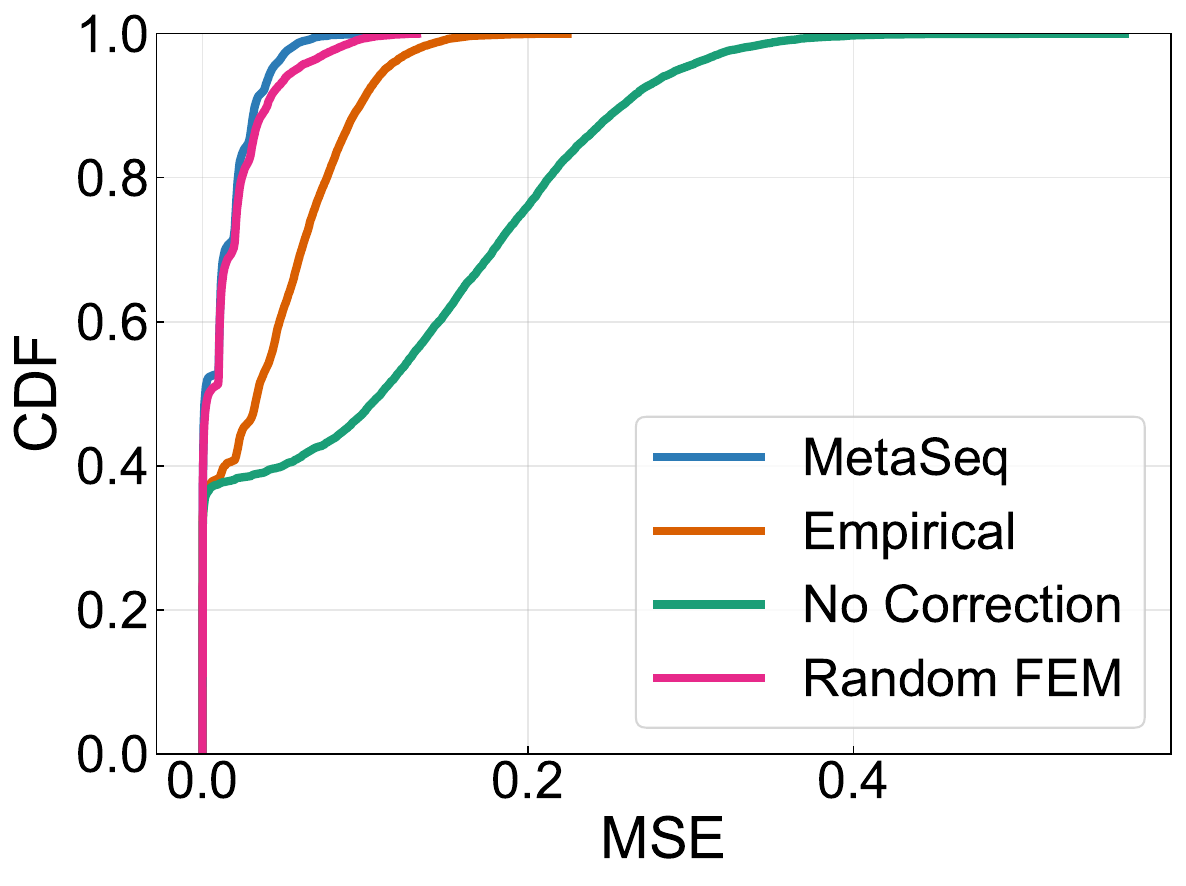}
		\label{fig:FEM_compare}
	}
	\vspace{-15pt}
	\caption{The comparison of MSE under different correction methods.} 
	\label{fig:Correction}
	\vspace{-10pt}
\end{figure}

We evaluate the fidelity of the response on a hold-out FEM test set. We quantify the Mean Squared Error (MSE) between the analytical predictions and the ground-truth FEM simulations using Eqn.~\ref{eqn:mse}. Here we evaluate two aspects of our proposed correction strategy:

\textit{(a) Required number of calibration samples.} Firstly, we evaluate the MSE and optimization time against the number of FEM calibration samples to determine the optimal size of the pilot calibration set $\mathcal{D}_{\text{calib}}$. For each experiment, we optimize the parameters on calibration set until convergence. Fig.~\ref{fig:MSE_time_FEM} shows that the MSE reaches a plateau beyond 40 samples. While the accuracy stabilizes, the total optimization time grows linearly. This indicates a remarkably small pilot set of 40 samples could achieve high physical fidelity with around 5 minutes optimization.   

\textit{(b) Comparison of correction methodologies.} We further compare the proposed length-stratified sampling strategy against baselines using the Cumulative Distribution Function (CDF) of the MSE, as illustrated in Fig.~\ref{fig:FEM_compare}. \sysname achieves the best average MSE of $0.011$, surpassing random sampling ($0.014$) through more representative coverage of diverse sequence lengths. Both the empirical parameters and no correction baselines exhibit significantly higher errors, with average MSE of $0.040$ and $0.116$, respectively. 

In conclusion, this efficient FEM-intervened correction ensures that the analytical solver retains both computational efficiency and high fidelity for rapid construction of a massive, topologically diverse dataset.

\section{Group-based Exploration and RL Policy Update}\label{app:grpo}
To stabilize the exploration of coupled hybrid heads, we employ Group Relative Policy Optimization (GRPO)~\cite{shao2024deepseekmath}. This group-based strategy fits our inverse design task for generating multiple solutions: evaluating a batch of candidates simultaneously naturally encourages the discovery of diverse equivalent structures. Specifically, for each target response, we decode a group of \emph{sampled sequences} rather than relying on a single deterministic baseline. During sampling, the discrete topology is explored via top-$k$ token truncation. Concurrently, continuous geometry is explored by injecting Gaussian noise into the sampled token's latent parameters: $\tilde{\mathbf{z}}_t=\mathbf{z}_t+\epsilon_t$, where $\epsilon_t \sim \mathcal{N}(\mathbf{0}, \sigma_{\mathrm{p}}^2 \mathbf{I})$. Crucially, applying this noise \emph{before} the token-specific bounded mapping ensures all explored dimensions naturally remain within physically valid ranges. Finally, the policy is updated using a group-normalized advantage $A_i = (R_i - \mu_R) / \sigma_R$, where $\mu_R$ and $\sigma_R$ are the mean and standard deviation of the rewards within the sampled group. This efficiently reinforces generated trajectories that outperform their peers. Practically, we set $k=2$ and $\sigma_{\mathrm{p}}=0.5$ to balance exploration diversity and optimization stability, which is evaluated in Sec.~\ref{subsec:params}.

\section{Diffusion Implementation}\label{app:diffu}

\begin{figure}[t]
	\centering
	\includegraphics[width=0.6\linewidth]{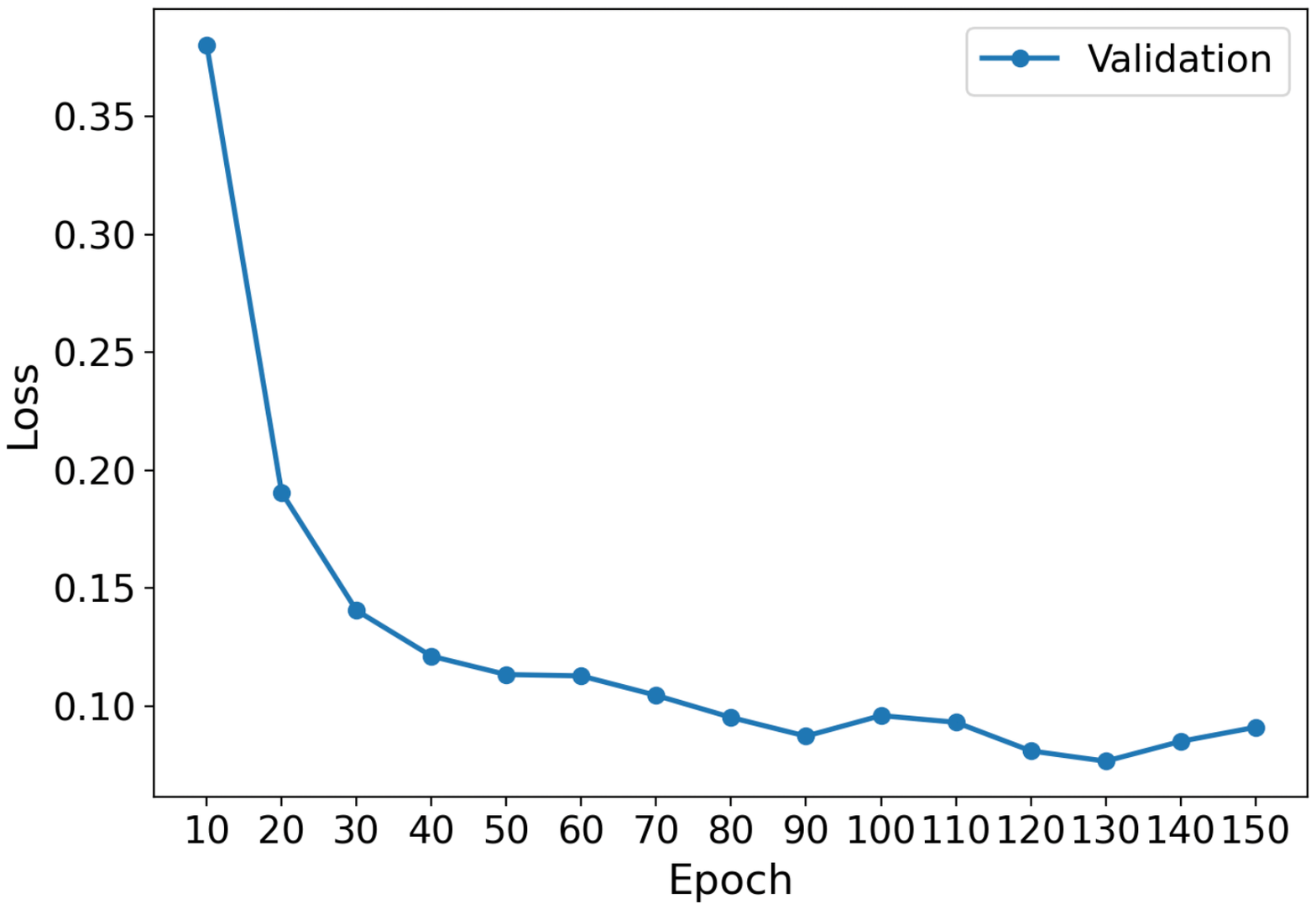}
	\vspace{-10pt}
	\caption{Validation loss across checkpoints.}
	\label{fig:validation_loss}
	\vspace{-20pt}
\end{figure}

\sssec{Architecture and Formulation.}
The generation module is designed to produce acoustic metamaterial patterns $\mathbf{x} \in \{0, 1\}^{96 \times 192}$ that correspond to a specific target acoustic response $\mathbf{y}$. This response is structured as a $100 \times 2$ matrix covering frequency-dependent transmission, and phase values, which is subsequently flattened into a $200$-dimensional condition vector. We formulate this task as a conditional image generation problem, employing a diffusion probabilistic model inspired by the MetaGen~\cite{zhao2025metagen} architecture. The diffusion process utilizes $T = 1000$ timesteps with a linear variance schedule ranging from $\beta_1 = 10^{-4}$ to $\beta_T = 0.02$. Noise prediction is handled by a conditional U-Net where both timestep and response embeddings are injected into the residual blocks to provide guidance during the denoising process.

\sssec{Data Preparation and Training.}
The initial dataset consists of $50,000$ simulated acoustic structures. During preprocessing, we removed $4,254$ cases to eliminate potential collisions caused by geometric compression. 
For the train-test split, our $7500$ samples dataset is filtered out from the initial dataset, and other remaining $38246$ samples were used for model training. Within the training partition, $1\%$ of the total dataset was designated for validation purposes. To ensure compatibility with the U-Net architecture, all structural geometries were converted into discretized binary images with dimensions of $96 \times 192$, which are divisible by $16$. 


\sssec{Evaluation and Simulation Pipeline.}
Training was conducted over $150$ epochs, with model checkpoints saved at $10$-epoch intervals. The selection of the best-performing model was based on validation performance, which involved converting the generated patterns into DXF files and running high-fidelity simulations in COMSOL. The primary metric for validation was the mean squared error (MSE) between the simulated and target acoustic responses. The validation loss across checkpoints is illustrated in Fig.~\ref{fig:validation_loss}.

Based on the validation performance, the 130-epoch checkpoint model was used to generate structures for the entire test dataset. These final designs were then simulated in COMSOL to verify the performance. The diffusion baseline achieved a final test MSE of 0.1055, representing the average error over the full test set.

\section{Forward Model}\label{app:forw}
\sssec{Forward surrogate model.}
To accelerate reward evaluation during RL fine-tuning, MetaSeq employs a fast, data-driven forward surrogate model as a forward evaluator. Given a generated MetaSeq design sequence, the model predicts its response over 100 frequency bins and serves as a GPU-efficient approximation to the calibrated physical solver. This allows response-space reward computation to be performed much more efficiently during training while preserving the same sequence-based design representation.

\sssec{Architecture.}
The forward surrogate model is implemented as a compact Transformer encoder operating on the token-parameter sequence. Each structural token is mapped to a token embedding, while its paired continuous geometric parameters are encoded into the same latent dimension and fused with the token representation by element-wise addition. The fused sequence is then processed by layer normalization, sinusoidal positional encoding, and a 4-layer Transformer encoder with hidden size $128$ and $4$ attention heads. To support variable-length designs, padding positions are masked during self-attention, and the final sequence representation is aggregated by masked mean pooling over valid tokens only. A prediction head then maps the pooled representation to the response vector, including transmission and phase across 100 frequency bins.

\sssec{Training objective.}
The forward surrogate model is trained on paired design-response data using a fixed train/validation/ test split with explicit leakage checking. Geometric parameters are normalized before being fed into the network, and the model is optimized with Adam and a OneCycle learning-rate schedule. The training loss is defined as the sum of mean squared error on transmission, together with a circular phase MSE (equal to Eq.~\ref{eq:rl_response_loss})  that accounts for phase periodicity. This formulation makes the model well aligned with MetaSeq's response-oriented optimization pipeline, where accurate and efficient response prediction is needed for large-scale RL evaluation.

\section{Microbenchmark: Hyperparameters}\label{app:hype}


\begin{figure}[t]
	\centering
	\subfigure[Fine-tune Scope.]{
		\centering
		\includegraphics[width=0.47\linewidth]{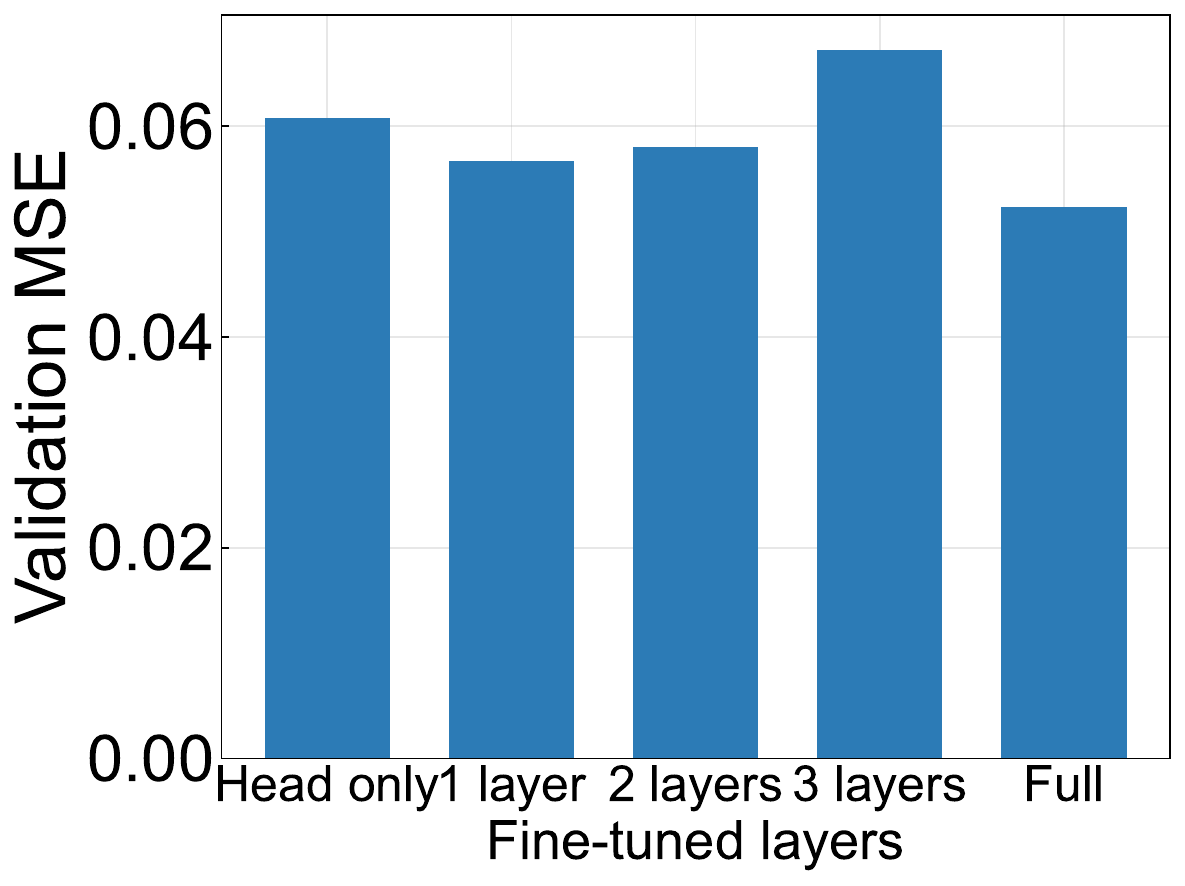}
		\label{fig:ft_scope}
	}\hfill
	\subfigure[Penalty Weights.]{
		\centering
		\includegraphics[width=0.47\linewidth]{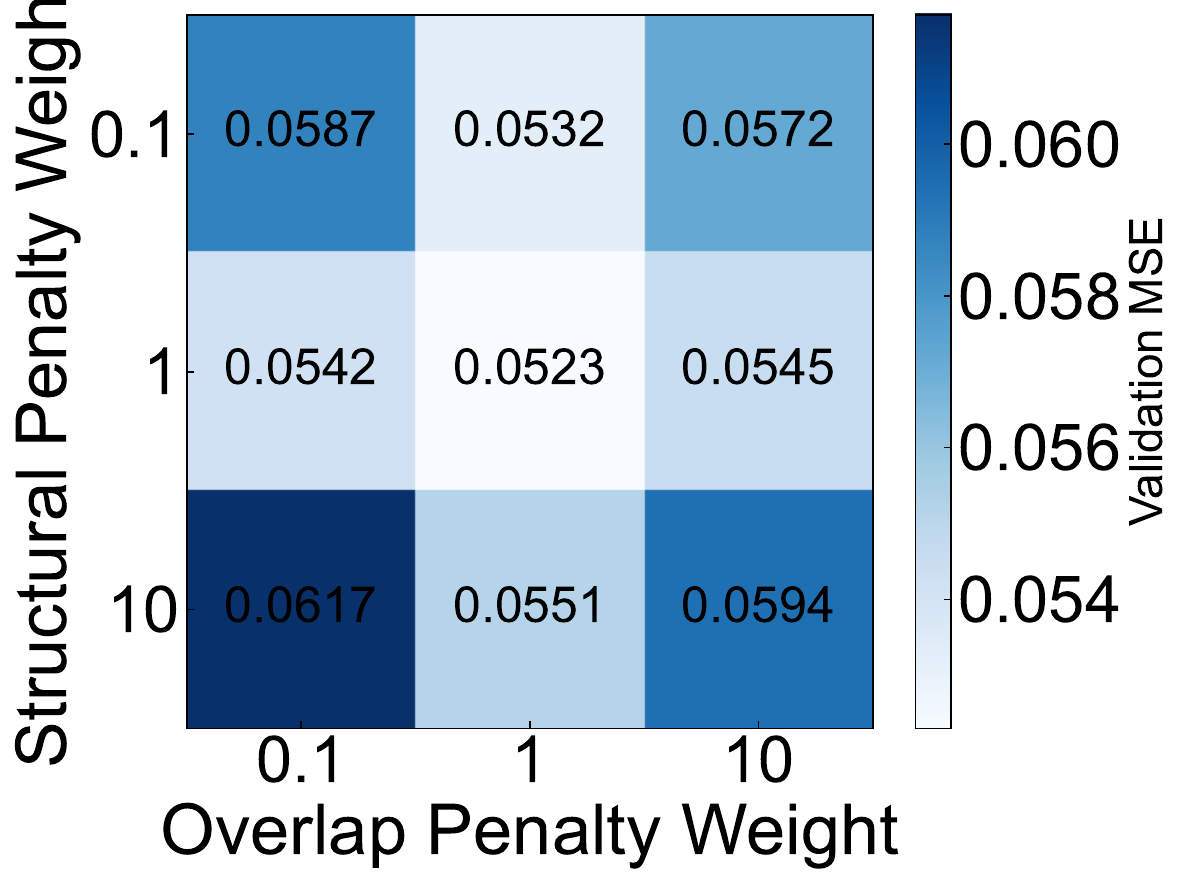}
		\label{fig:weight}
	}
	
	\subfigure[Top-k Sampling.]{
		\centering
		\includegraphics[width=0.47\linewidth]{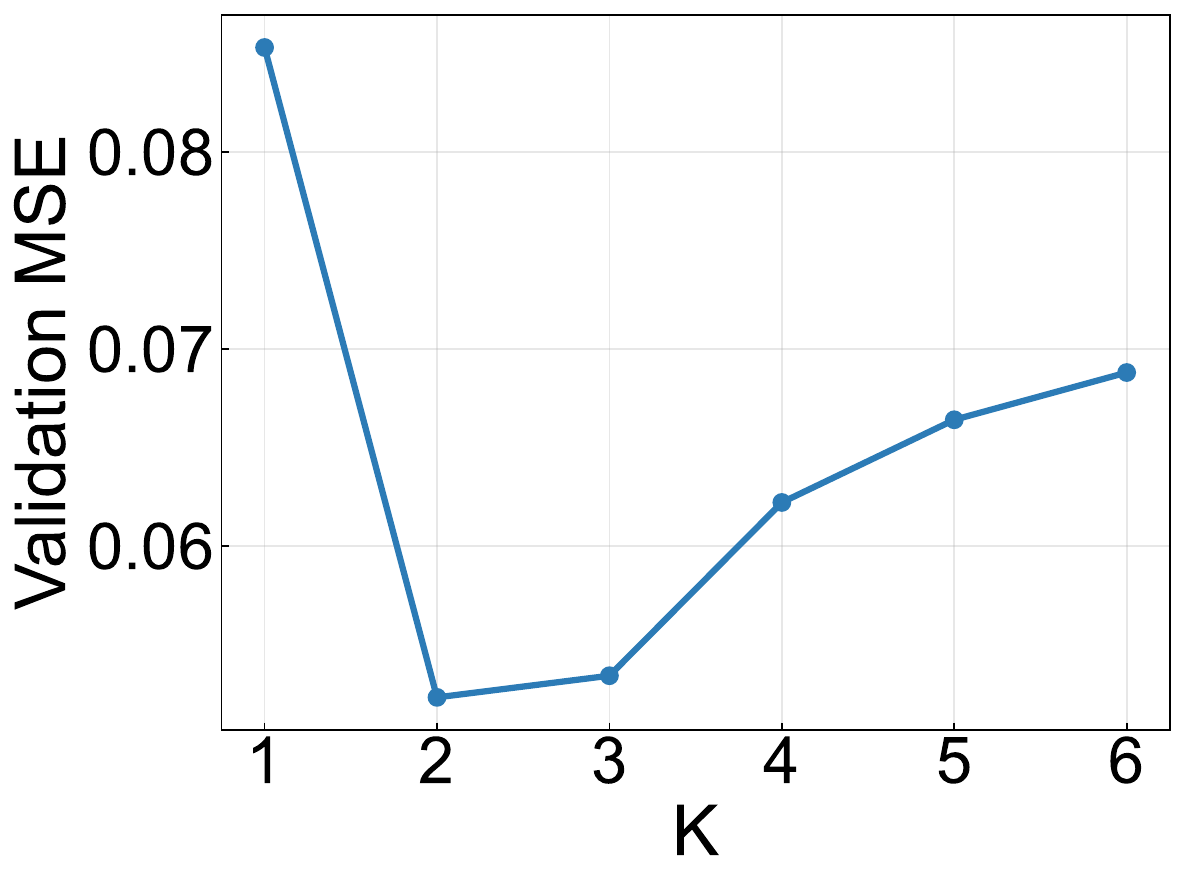}
		\label{fig:topk}
	}\hfill
	\subfigure[Parameter Deviation.]{
		\centering
		\includegraphics[width=0.47\linewidth]{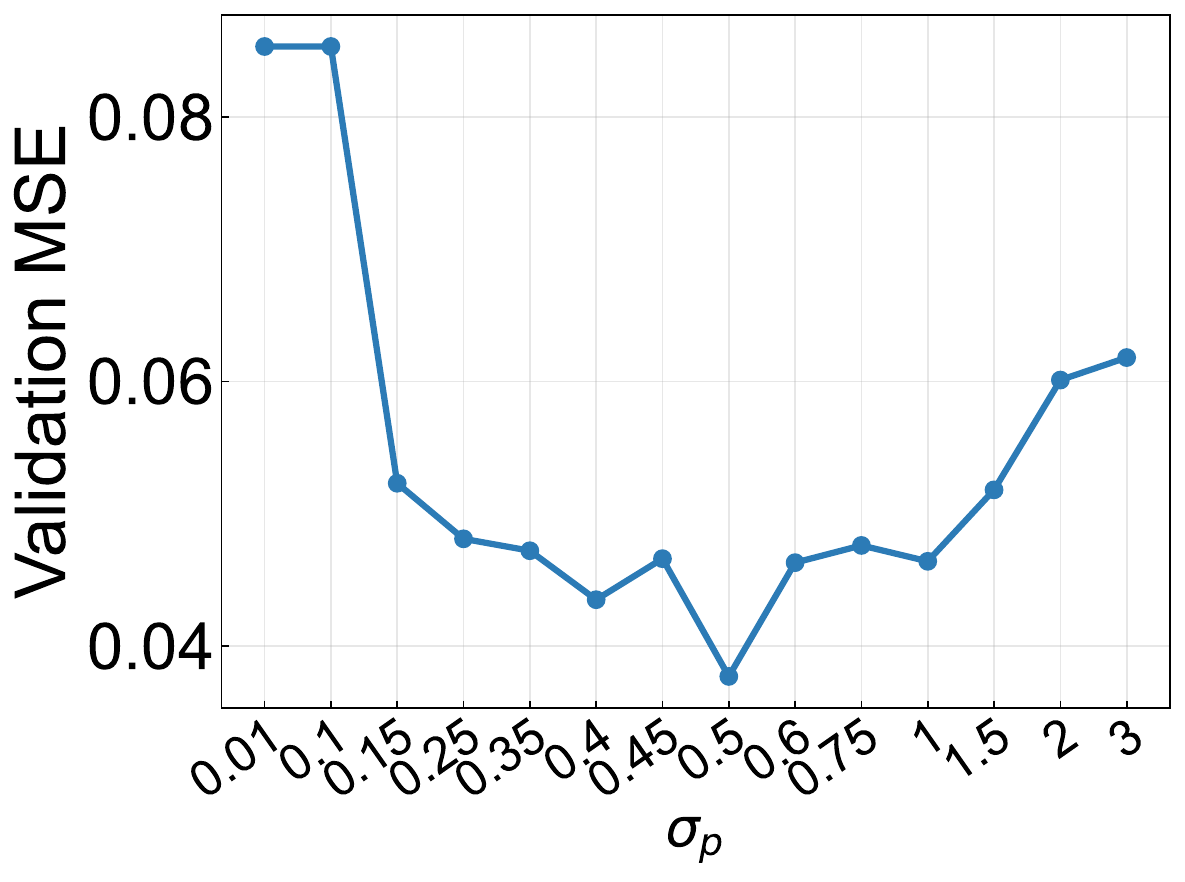}
		\label{fig:param_std}
	}
	\vspace{-15pt}
	\caption{Hyperparameters selection during training.} 
	\label{fig:micro}
	\vspace{-15pt}
\end{figure}

\sssec{Fine-Tune Scope.} We compare three fine-tuning scopes: (1) output heads only, (2) output heads plus the top decoder layers, and (3) the full model. As shown in Fig.~\ref{fig:ft_scope}, full-model fine-tuning performs the best.
We therefore use full-model fine-tuning in all subsequent experiments.

\sssec{Penalty Weights.} The RL reward incorporates two penalties—structural invalidity and geometric overlap—to balance performance optimization and constraint satisfaction. We sweep  both coefficients (e.g., structural penalty weight and overlap penalty weight). Fig.~\ref{fig:weight} shows that both penalties are equally important, so we employ equal weights.


%

%

\sssec{Top-k Sampling.} $k$ regulates exploration over discrete topology tokens. While the one-to-many nature of inverse design requires exploring alternative structures, excessive stochasticity can destabilize generation. Fig.~\ref{fig:topk} shows that limited exploration yields the best results: nearly greedy sampling is overly restrictive, while larger $k$ values degrade performance. We therefore set $k=2$.

\sssec{Parameter Deviation.} For continuous geometry parameters, RL explores via Gaussian perturbations in the latent parameter space. The standard deviation $\sigma_p$ governs the exploration scale: insufficient values may trap the policy near the pretrained solution, while excessive noise destabilizes the search process. As illustrated in Fig.~\ref{fig:param_std}, performance peaks at a moderate  exploration  level and degrades as $\sigma_p$ becomes overly aggressive. We therefore adopt $\sigma_p = 0.5$ in all experiments.

%

\section{Microbenchmark: balanced dataset} \label{app:ban_dataset}

\begin{figure}[t]
	\centering
	\includegraphics[width=0.5\linewidth]{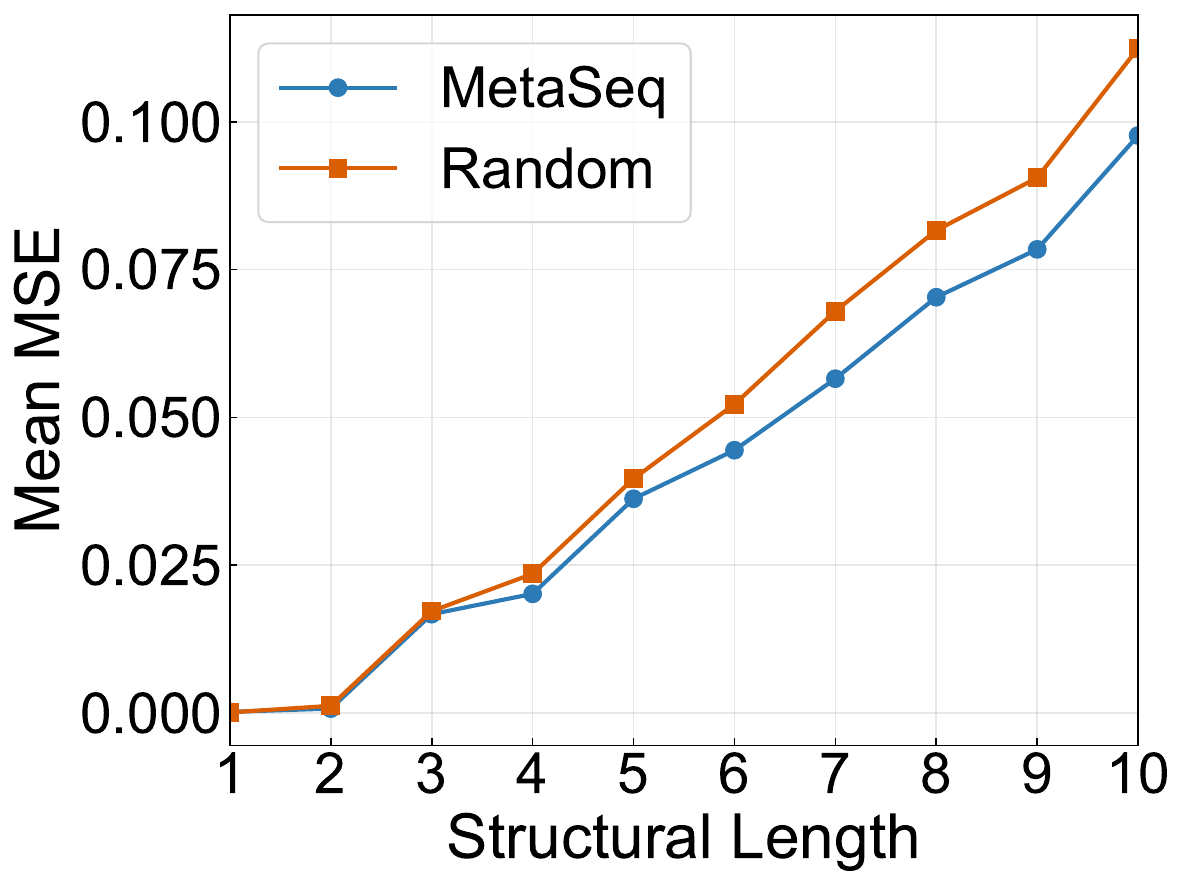}
	\vspace{-10pt}
	\caption{Performance under balanced dataset and random dataset.} 
	\label{fig:balanced_vs_unbalanced}
	\vspace{-15pt}
\end{figure}
We next evaluate our balancing strategy in Sec.~\ref{sec:dataset}. We train identical models  on  the balanced dataset and on a random generated baseline, and evaluate on the same test set. As shown in Fig.~\ref{fig:balanced_vs_unbalanced}, the balanced dataset achieves better accuracy in mean MSE, and the gap widens as structural complexity increases.

\end{document}